\documentclass[preprint,12pt]{elsarticle}
\usepackage{amsmath}
\usepackage{amssymb}
\usepackage[utf8]{inputenc}
\usepackage{caption}
\usepackage{graphicx}
\usepackage{bbding}
\usepackage{caption}
\usepackage{subcaption}
\usepackage{hyperref}
\usepackage{multirow}
\usepackage{adjustbox}
\usepackage{tikz}

\usepackage{acro}

\DeclareAcronym{lrp}{
  short = LRP ,
  long  = Layer-wise Relevance Propagation ,
  tag = abbrev
}

\DeclareAcronym{dl}{
  short = DL ,
  long  = deep learning ,
  tag = abbrev
}

\DeclareAcronym{ml}{
  short = ML ,
  long  = machine learning ,
  tag = abbrev
}

\DeclareAcronym{gan}{
  short = GAN ,
  long  = generative adversarial network ,
  tag = abbrev
}

\DeclareAcronym{idl}{
  short = iDL ,
  long  = interpretable deep learning ,
  tag = abbrev
}

\DeclareAcronym{xai}{
  short = XAI ,
  long  = explainable artificial intelligence ,
  tag = abbrev
}

\DeclareAcronym{gdpr}{
  short = GDPR ,
  long  = General Data Protection Regulation ,
  tag = abbrev
}

\DeclareAcronym{hiv}{
  short = HIV ,
  long  = Human Immunodeficiency Virus ,
  tag = abbrev
}

\DeclareAcronym{cnn}{
  short = CNN ,
  long  = convolutional neural network ,
  tag = abbrev
}

\DeclareAcronym{ae}{
  short = AE ,
  long  = autoencoder ,
  tag = abbrev
}

\DeclareAcronym{vae}{
  short = VAE ,
  long  = variational autoencoder ,
  tag = abbrev
}

\DeclareAcronym{nn}{
  short = NN ,
  long  = neural network ,
  tag = abbrev
}

\DeclareAcronym{rnn}{
  short = RNN ,
  long  = recurrent neural network ,
  tag = abbrev
}

\DeclareAcronym{crnn}{
  short = C-RNN ,
  long  = convolutional recurrent neural network ,
  tag = abbrev
}

\DeclareAcronym{mlp}{
  short = MLP ,
  long  = multi-layer perceptron ,
  tag = abbrev
}

\DeclareAcronym{dnn}{
  short = DNN ,
  long  = deep neural network ,
  tag = abbrev
}

\DeclareAcronym{gap}{
  short = GAP ,
  long  = global average pooling ,
  tag = abbrev
}

\DeclareAcronym{svm}{
  short = SVM ,
  long  = support vector machine ,
  tag = abbrev
}

\DeclareAcronym{smri}{
  short = sMRI ,
  long  = structural magnetic resonance imaging ,
  tag = abbrev
}

\DeclareAcronym{ic}{
  short = IC ,
  long  = independent component ,
  tag = abbrev
}

\DeclareAcronym{fbn}{
  short = FBN ,
  long  = functional brain networks ,
  tag = abbrev
}

\DeclareAcronym{fmri}{
  short = fMRI ,
  long  = functional magnetic resonance imaging ,
  tag = abbrev
}

\DeclareAcronym{rsfmri}{
  short = rs-fMRI ,
  long  = resting-state functional magnetic resonance imaging ,
  tag = abbrev
}

\DeclareAcronym{mri}{
  short = MRI ,
  long  = magnetic resonance imaging ,
  tag = abbrev
}

\DeclareAcronym{dti}{
  short = DTI ,
  long = diffusion tensor imaging ,
  tag = abbrev
}

\DeclareAcronym{srddl}{
  short = sr-DDL ,
  long = structurally-regularized Dynamic Dictionary Learning ,
  tag = abbrev
}

\DeclareAcronym{mmse}{
  short = MMSE ,
  long = mini-mental state examination ,
  tag = abbrev
}

\DeclareAcronym{scm}{
  short = SCM ,
  long = Structural Causal Model ,
  tag = abbrev
}

\DeclareAcronym{dscm}{
  short = DSCM ,
  long = Deep Structural Causal Model ,
  tag = abbrev
}

\DeclareAcronym{fnc}{
  short = FNC ,
  long  = functional network connectivity ,
  tag = abbrev
}

\DeclareAcronym{tc}{
  short = TCs ,
  long  = time courses ,
  tag = abbrev
}

\DeclareAcronym{cn}{
  short = CN ,
  long  = cognitively normal ,
  tag = abbrev
}

\DeclareAcronym{ad}{
  short = AD ,
  long  = Alzheimer's Disease ,
  tag = abbrev
}

\DeclareAcronym{adhd}{
  short = ADHD ,
  long = attention deficit hyperactivity disorder ,
  tag = abbrev
}

\DeclareAcronym{mci}{
  short = MCI ,
  long  = mild cognitive impairment ,
  tag = abbrev
}

\DeclareAcronym{pmci}{
  short = pMCI ,
  long  = progressive mild cognitive impairment ,
  tag = abbrev
}

\DeclareAcronym{smci}{
  short = sMCI ,
  long  = stable mild cognitive impairment ,
  tag = abbrev
}

\DeclareAcronym{adni}{
  short = ADNI ,
  long  = Alzheimer's Disease Neuroimaging Initiative ,
  tag = abbrev
}
\DeclareAcronym{aibl}{
  short = AIBL,
  long  = Australian Imaging Biomarkers and Lifestyle Study of Ageing,
  tag = abbrev
}

\DeclareAcronym{gnn}{
short=GNN,
long=graph neural network,
tag=abbrev
}

\DeclareAcronym{gcn}{
short=GCN,
long=graph convolutional network,
tag=abbrev
}

\DeclareAcronym{asd}{
short=ASD,
long=Autism Spectrum Disorders,
tag=abbrev
}

\DeclareAcronym{pd}{
short=PD,
long=Parkinson's Disease,
tag=abbrev
}

\DeclareAcronym{scz}{
short=SCZ,
long=Schizophrenia,
tag=abbrev
}

\DeclareAcronym{hc}{
short=HC,
long= healthy control ,
tag=abbrev
}

\DeclareAcronym{ms}{
short=MS,
long=Multiple Sclerosis,
tag=abbrev
}

\DeclareAcronym{lstm}{
short=LSTM,
long=long short-term memory,
tag=abbrev
}

\DeclareAcronym{relu}{
short=ReLU,
long=Rectified Linear Unit,
tag=abbrev
}

\DeclareAcronym{nlp}{
short=NLP,
long=natural language processing,
tag=abbrev
}

\DeclareAcronym{ct}{
short=CT,
long=Computed Tomography,
tag=abbrev
}

\DeclareAcronym{brats}{
short=BraTS,
long=Brain Tumor Segmentation,
tag=abbrev
}

\DeclareAcronym{ich}{
short=ICH,
long=intracerebral hemorrhage,
tag=abbrev
}

\DeclareAcronym{spect}{
short=SPECT,
long=Single Photon Emission
Computed Tomography,
tag=abbrev
}

\DeclareAcronym{abide}{
short=ABIDE,
long=Autism Brain Imaging Data Exchange,
tag=abbrev
}

\DeclareAcronym{ncc}{
short=NCC,
long=normalised cross correlation,
tag=abbrev
}

\DeclareAcronym{hcp}{
short=HCP,
long=Human Connectome Project,
tag=abbrev
}

\DeclareAcronym{dhcp}{
short=dHCP,
long=developing Human Connectome Project,
tag=abbrev
}

\DeclareAcronym{kki}{
short=KKI,
long=Kennedy Krieger Institute,
tag=abbrev
}

\DeclareAcronym{wrat}{
short=WRAT,
long=Wide Range Achievement Test,
tag=abbrev
}

\DeclareAcronym{pnc}{
short=PNC,
long=Philadelphia Neurodevelopmental Cohort,
tag=abbrev
}

\DeclareAcronym{hande}{
short=H\&E,
long=hematoxylin \& eosin,
tag=abbrev
}

\DeclareAcronym{oasis}{
short=OASIS,
long=Open Access Series of Imaging Studies,
tag=abbrev
}

\DeclareAcronym{fdgpet}{
short=FDG-PET,
long=fluorodeoxyglucose positron emission tomography,
tag=abbrev
}

\DeclareAcronym{fc}{
short=FC,
long=functional connectivity,
tag=abbrev
}

\DeclareAcronym{us}{
short=US,
long=ultrasound,
tag=abbrev
}

\DeclareAcronym{ftd}{
short=FTD,
long=Frontotemporal Dementia,
tag=abbrev
}

\DeclareAcronym{ddpm}{
short=DDPM,
long=Denoising Diffusion Probabilistic Model,
tag=abbrev
}

\DeclareAcronym{ppmi}{
short=PPMI,
long=Parkinson’s Progression Markers Initiative,
tag=abbrev
}

\newcommand{\etal}[1]{\textit{et al.}}
\newcommand{\myetal}[1]{#1 \textit{et al.}}

\begin{document}

\begin{frontmatter}
\title{Applications of interpretable deep learning in neuroimaging: a comprehensive review}

\author[inst1]{Lindsay Munroe\corref{cor1}}
\affiliation[inst1]{organization={Department of Neuroimaging},
            addressline={King's College London}, 
            city={London},
            country={UK}}

\author[inst2]{Mariana da Silva}
\affiliation[inst2]{organization={School of Biomedical Engineering and Imaging Sciences},
            addressline={King's College London}, 
            city={London},
            country={UK}}

\author[inst3]{Faezeh Heidari}
\affiliation[inst3]{organization={Institute of Clinical Medicine},
            addressline={University of Eastern Finland}, 
            city={Kuopio},
            country={Finland}}

\author[inst2]{Irina Grigorescu}
\author[inst2]{Simon Dahan}
\author[inst2]{Emma C. Robinson}
\author[inst2]{Maria Deprez}
\author[inst1]{Po-Wah So}

\cortext[cor1]{Corresponding author: lindsay.munroe@kcl.ac.uk}

\begin{abstract}

Clinical adoption of deep learning models has been hindered, in part, because the ``black-box" nature of neural networks leads to concerns regarding their trustworthiness and reliability. These concerns are particularly relevant in the field of neuroimaging due to the complex brain phenotypes and inter-subject heterogeneity often encountered. The challenge can be addressed by interpretable deep learning (iDL) methods that enable the visualisation and interpretation of the inner workings of deep learning models. This study systematically reviewed the literature on neuroimaging applications of iDL methods and critically analysed how iDL explanation properties were evaluated. Seventy-five studies were included, and ten categories of iDL methods were identified. We also reviewed five properties of iDL explanations that were analysed in the included studies: biological validity, robustness, continuity, selectivity, and downstream task performance. We found that the most popular iDL approaches used in the literature may be sub-optimal for neuroimaging data, and we discussed possible future directions for the field. \\

\end{abstract}

\end{frontmatter}
\clearpage

\section{Introduction}\label{sec:introduction}

Traditionally, analysis and interpretation of neuroimaging data requires specialised expertise, is often laborious, and is subject to inter-observer variability. Therefore, \ac{dl} has become a popular tool in neuroimaging in recent years, driven by the rise in computer processing power as well as increased access to large medical imaging datasets and the success of novel model architectures. In neuroimaging, \ac{dl} has been applied to segmentation \cite{de2015deep,chen2018voxresnet,milletari2017hough}, super-resolution \cite{kang2015prediction,xu2020ultra}, image synthesis \cite{gong2018deep,kwon2019generation,shin2018medical} and classification \cite{liu2018landmark, bohle2019layer}, among other applications. Despite the success of \ac{dl} for analysing and interpreting neuroimaging data, adoption remains limited partly because \ac{dl} models are often opaque and considered to be ``black boxes". In other words, the internal workings of \ac{dl} models are not comprehensible to humans, which leads to concerns regarding their reliability and trustworthiness. Indeed, such ``black box" models do not satisfy European \ac{gdpr} legal requirements to provide ``information about the logic involved" \cite{hacker2020explainable}.\\

\subsection{Advantages of interpretable deep learning}
\Ac{idl} has been proposed to address the opacity problem of \ac{dl} models, for example, by producing explanations that highlight brain regions that are most relevant for the model predictions. \ac{idl} methods can support the translation of \ac{dl} to the clinic by providing healthcare practitioners with explanations to verify predictions and communicate with patients. Additionally, deep learning practitioners can leverage \ac{idl} to debug their models and identify cases where a model makes the right decision for the wrong reason \cite{lapuschkin2019unmasking}. \ac{idl} methods can also be employed to test scientific hypotheses, such as identifying brain regions involved in disease pathogenesis. \\

\subsection{Evaluation of iDL explanations}
A challenging aspect of \ac{idl} is assessing the quality of explanations because such ground truths are typically unavailable. While experts such as clinicians, pathologists, or imaging scientists can qualitatively evaluate explanations, quantitative and automated metrics are often preferred, particularly when access to medical professionals is limited. Researchers have proposed various quantitative methods to evaluate desirable properties of \ac{idl} explanations, with a particular focus on assessing fidelity and robustness (\textit{e.g.}, \cite{adebayo2018sanity}, \cite{montavon2018methods}, \cite{kindermans2019reliability}, \cite{hooker2019benchmark}, \cite{samek2016evaluating}, \cite{yang2019benchmarking}, \cite{lapuschkin2016analyzing}). \\

Fidelity refers to the extent to which explanations reflect the inner workings of the associated deep learning model. Fidelity is usually evaluated by removing features or comparing the explanations to ground truth, if available. In computer vision, feature-removal approaches generally involve masking image regions with the highest relevance in the associated explanation, obtaining predictions for the modified images, and then measuring the change in model output or accuracy. A substantial drop in accuracy indicates that the explanations faithfully highlight image features attended to by the model. For example, Montavon \etal{} developed a procedure to assess fidelity in which they iteratively removed $4 \times 4$ patches from images with the highest relevance and plotted the number of patches removed against model output score  \cite{montavon2018methods}. In another example of fidelity evaluation, Adebayo \etal{} randomised model parameters and data labels as two sanity checks to assess whether \ac{idl} explanations truly reflect either the model mechanisms or the relationship between image features and the label \cite{adebayo2018sanity}. Alternatively, explanations can be compared to ground truth maps of image features the model is expected to attend to when making predictions. For instance, bounding box annotations for objects in natural images have been used as ground truth and the ratio of mean relevance outside \textit{vs.} inside the bounding box has been calculated to assess the fidelity of explanations \cite{lapuschkin2016analyzing}. \\

Robustness can be described as the stability of model explanations under varying modelling conditions. For example, Montavon \etal{} introduced the concept of continuity, which means that an \ac{idl} method should produce similar explanations for similar input images \cite{montavon2018methods}. The evaluation of \ac{idl} methods is an active research field, and for a comprehensive review of the topic, we recommend readers to refer to Alangari \etal{}'s work \cite{alangari2023exploring}. \\

\subsection{Classification of interpretable deep learning methods}
Two main categories of \ac{idl} methods exist: post-hoc and intrinsic. \textit{Post-hoc} methods use reverse engineering to generate an explanation from a ``black-box" model after training. In contrast, \textit{intrinsic} methods incorporate interpretable components into the model architecture during the design phase. Another way to classify interpretable methods is by local vs. global explanations. \textit{Local} explanations focus on individual samples and thereby increase trust in the model outcomes, whereas global explanations seek to provide a deeper understanding of the mechanism by which the model works. \\

\subsection{Study objectives}

The objectives of this review are: 

\begin{enumerate} 
    \item To systematically review \ac{idl} methods applied to neuroimaging studies. 

    \item To review the evaluation of \ac{idl} explanations in the studies included in this review, explicitly identifying the properties evaluated and associated quantitative metrics proposed. 
\end{enumerate} 

To the best of our knowledge, this is the first study to systematically review both post-hoc and intrinsic \ac{idl} methods in the field of neuroimaging. \\

We have further sub-classified \ac{idl} methods of the two categories (Table \ref{tab:all_methods_v3}). Initially, we introduce five post-hoc methods (Section~\ref{sec:posthoc_methods}) and five intrinsic methods \ref{sec:intrinsic_methods} before reviewing applications to neuroimaging for each method (Section~\ref{sec:applications}). Finally, we consider how \ac{idl} explanations were evaluated across the included studies (Section~\ref{sec:evaluationofxai}).\\

\begin{figure}
    \centering
    \includegraphics[width=14cm]{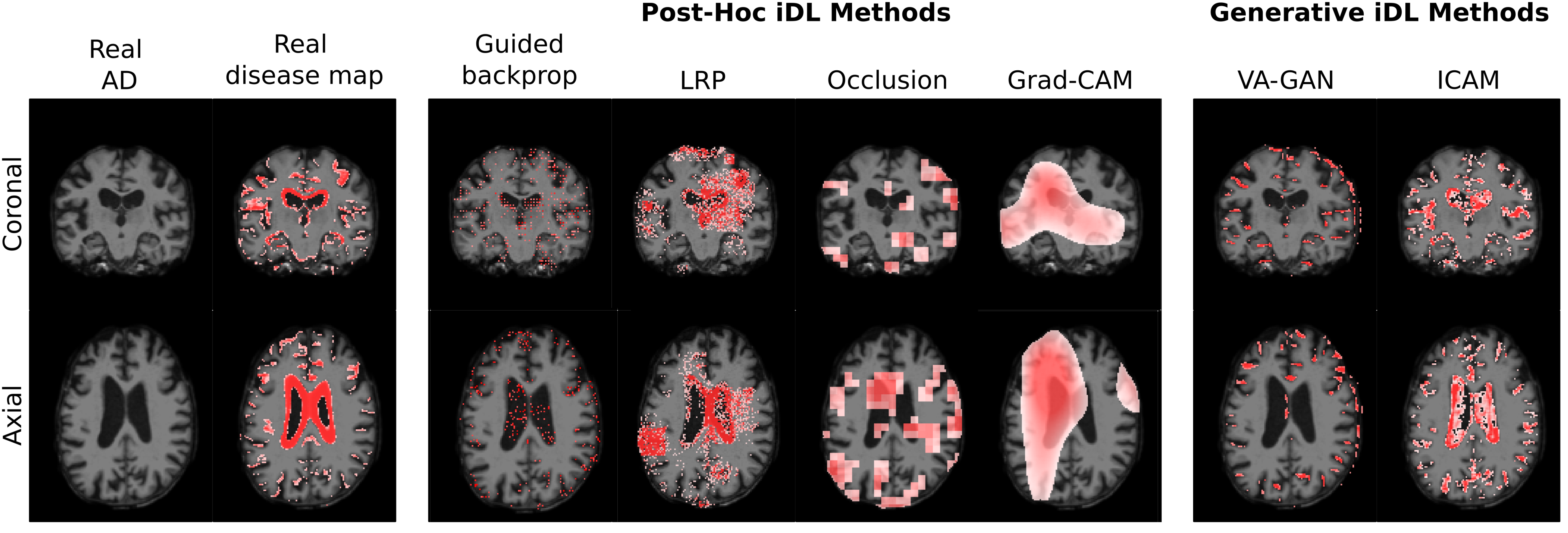}
    \caption{\textbf{} Comparison of post-hoc interpretability maps and generative interpretability methods applied to the classification of Alzheimer's disease (AD) vs Mild cognitive impairment (MCI) in brain MRI volumes. The real disease map is the ``ground-truth'' shown for comparison. Figure adapted from Bass \textit{et al.} \cite{bass2022icam}}
    \label{adni_comparison}
\end{figure}

 \section{Systematic search methodology}\label{sec:appendix_search_methodology}

We identified relevant articles for this review by querying PubMed, Web of Science, Google Scholar and arXiv using the following search terms: 1. explainable, 2. XAI, 3. interpretable, 4. explainability, 5. interpretability, 6. causal reasoning, 7. counterfactuals, 8. deep learning, 9. AI, 10. neural network, 11. machine learning, 12. brain imaging, 13. neuroimaging, 14. neuroradiology. The search terms were combined in the logical statement (1 OR 2 OR 3 OR 4 OR 5 OR 6 OR 7) AND (8 OR 9 OR 10 OR 11) AND (12 OR 13 OR 14). Articles from 2015 were included for PubMed and Google Scholar, whereas all years were included for Web of Science and arXiV due to the small number of articles returned. \\


Articles were initially screened based on the article title and abstract and accepted or rejected from a full-text review based on inclusion and exclusion criteria (Table \ref{tab:screening_criteria}). Only the first 500 results from Google Scholar were screened because later results were irrelevant. Finally, we extracted the pertinent information from all accepted articles into a spreadsheet for further analysis.\\

\begin{table}[]
    \centering
    \tabcolsep=0.11cm
    \begin{tabular}{ p{12cm} }
        \hline
        \textbf{Inclusion/ exclusion criteria for article screening} \\ 
        \hline
        \textbf{Include}...both \emph{in-vivo} and \emph{ex-vivo} imaging.  \\
        \textbf{Exclude}...non-human subjects. \\
        \textbf{Include}...the following imaging modalities: structural and functional magnetic resonance imaging, computed tomography, and positron emission tomography. \\
        \textbf{Exclude}...electroencephalogram and magnetoencephalography data. \\
        \textbf{Exclude}...non-peer reviewed articles.\\
        \textbf{Exclude}...non-English language articles. \\
        \textbf{Exclude}...PhD and Masters theses. \\
        \textbf{Exclude}...reviews, surveys, opinion articles and books. Articles must implement at least one interpretable deep learning method. \\
        \textbf{Exclude}...interpretable methods applied to machine learning models other than neural networks. \\
        \textbf{Exclude}...for quality control. For example, if the explanations could not be reasonably interpreted. \\
        \hline
    \end{tabular}
    \caption{Inclusion and exclusion criteria}
    \label{tab:screening_criteria}
\end{table}

\section{Overview}

\begin{figure}
    \centering
    \includegraphics[width=1.0\textwidth]{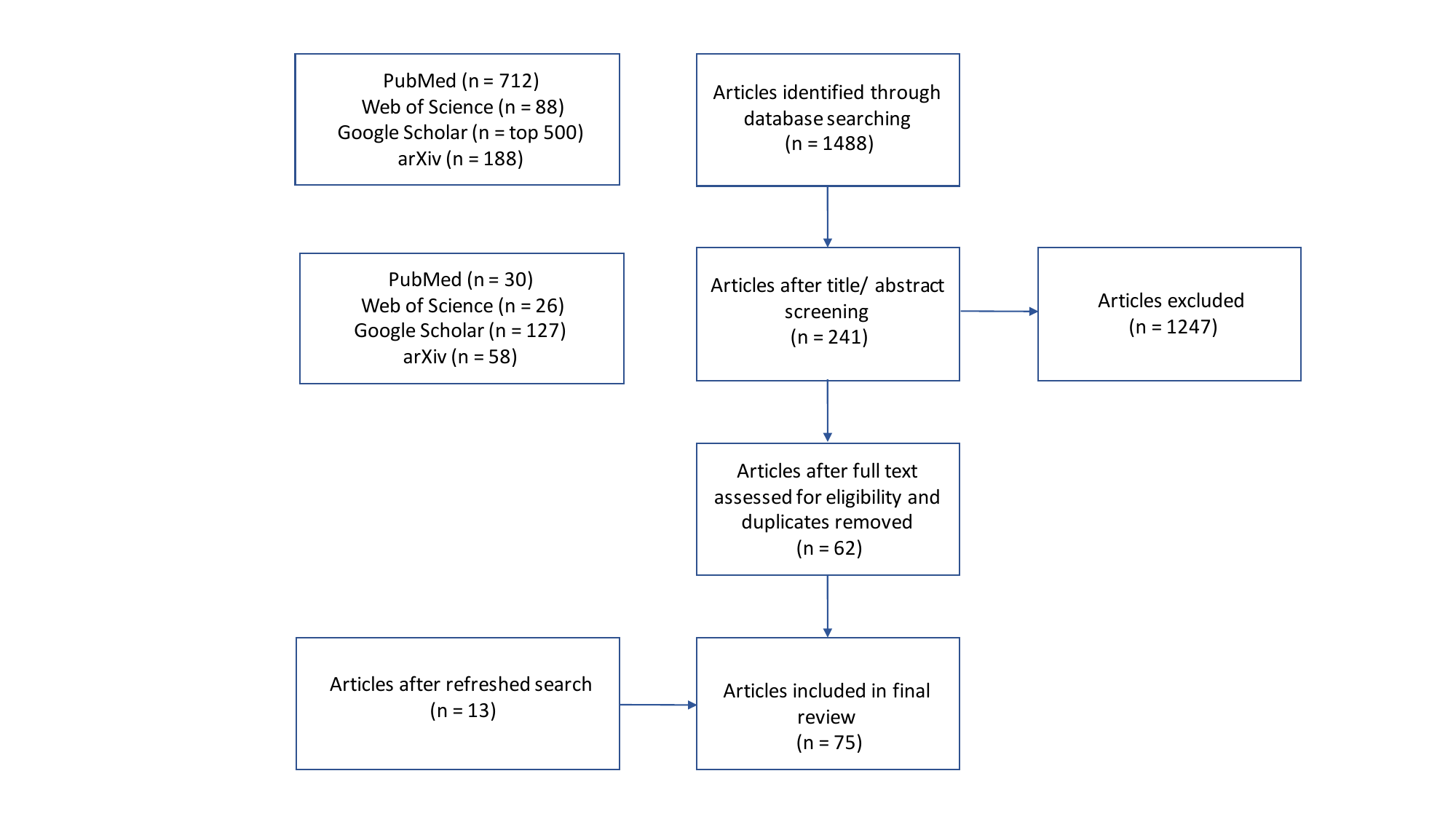} 
    \caption{The Preferred Reporting Items for Systematic Reviews and Meta-Analyses (PRISMA) flowchart}
    \label{fig:systematic_review_flowchart}
\end{figure}

The number of articles returned was 712 for PubMed, 88 for Web of Science, 1000 for Google Scholar (the upper limit), and 189 for arXiV. After title and abstract screening, the number of accepted articles was 30 for PubMed, 26 for Web of Science, 127 for Google Scholar and 58 for arXiV. After full-text review and removal of duplicates, and added articles after a refresh, the number of accepted articles was 75 (Fig. \ref{fig:systematic_review_flowchart}). Table~\ref{tab:all_methods_v3} summarises the methods and papers introduced in this review. 

\begin{table}[h]
\footnotesize
\centering
\tabcolsep=0.11cm
\begin{tabular}{cll}
\hline  
\multicolumn{2}{c}{\textbf{Methods}} & \multicolumn{1}{c}{\textbf{Papers}} \\
\hline \hline \\
\multirow{22}{*}{\rotatebox[origin=c]{90}{\textbf{Post-hoc methods}}} & \multicolumn{2}{l}{ Tab.~\ref{tab:perturbation_based_methods} \textbf{Perturbation-based methods}} \\
 & \phantom{Tab..10}$\blacktriangleright$ Disease Classification & \cite{eitel2019testing}, \cite{liu2019deep}, \cite{yang2018visual}, \cite{nigri2020explainable}, \cite{thibeau2020visualization}, \cite{shahamat2020brain},  \cite{tang2019interpretable} \\
 & & \cite{magesh2020explainable}, \cite{li2018brain}, \cite{dhurandhar2018explanations}, \cite{shahamat2020brain}, \cite{mellema2020architectural}, \cite{yan2019discriminating} \\
 & \phantom{Tab..10}$\blacktriangleright$ Sex Classification & \cite{kan2020interpretation} \\
 & \phantom{Tab..10}$\blacktriangleright$ Brain Age Regression & \cite{bintsi2021voxel} \\
 & \multicolumn{2}{l}{ Tab.~\ref{tab:gradient_based_methods} \textbf{Gradient-based methods} } \\
  & \phantom{Tab..10}$\blacktriangleright$ Disease Classification & \cite{oh2019classification}, \cite{eitel2019testing}, \cite{essemlali2020understanding},  \cite{li2019graph}, \cite{zhao2019confounder}\\
  & \phantom{Tab..10}$\blacktriangleright$ Brain Age Regression & \cite{levakov2020deep} \\
  & \phantom{Tab..10}$\blacktriangleright$ Cognitive Task Decoding & \cite{mcclure2020evaluating}, \cite{ismail2019input} \\
& \multicolumn{2}{l}{ Tab.~\ref{tab:backpropagation_based_methods} \textbf{Backpropagation-based methods} } \\
 & \phantom{Tab..10}$\blacktriangleright$ Disease Classification & \cite{bohle2019layer}, \cite{eitel2019testing} \\
 & \phantom{Tab..10}$\blacktriangleright$ Sex Classification & \cite{kan2020interpretation} \\
 & \phantom{Tab..10}$\blacktriangleright$ Cognitive Task Decoding & \cite{thomas2019analyzing} \\
& \multicolumn{2}{l}{ Tab.~\ref{tab:cam_methods} \textbf{Class activation maps}} \\
  & \phantom{Tab..10}$\blacktriangleright$ Disease Classification & \cite{zhang2021explainable}, \cite{yang2018visual}, \cite{azcona2020interpretation}, \cite{khan2019transfer}, \cite{tang2019interpretable}, \cite{williamson2022improving}, \cite{li2020pooling}, \cite{windisch2020implementation}, \cite{lee2019explainable} \\
 & \phantom{Tab..10}$\blacktriangleright$ Sex Classification & \cite{gao2019dense}, \cite{kim2020understanding}, \cite{kan2020interpretation} \\
& \phantom{Tab..10}$\blacktriangleright$ Tissue Segmentation & \cite{natekar2020demystifying} \\
 & \phantom{Tab..10}$\blacktriangleright$ Cognitive Score Prediction & \cite{hu2021interpretable}, \cite{qu2021ensemble} \\
& \multicolumn{2}{l}{ Tab.~\ref{tab:weight_analysis_methods} \textbf{Weight Analysis} } \\
  & \phantom{Tab..10}$\blacktriangleright$ Disease Classification & \cite{dvornek2019jointly}, \cite{li2021braingnn} \\
  & \phantom{Tab..10}$\blacktriangleright$ Tissue Segmentation & \cite{kori2020abstracting}, \cite{natekar2020demystifying} \\
  & \phantom{Tab..10}$\blacktriangleright$ Cognitive Task Decoding & \cite{li2021braingnn} \\ \\ \hline \\ 
\multirow{15}{*}{\rotatebox[origin=c]{90}{\textbf{Intrinsic methods}}} & \multicolumn{2}{l}{ Tab.~\ref{tab:disentangled_latent_space_methods} \textbf{Disentangled latent spaces} } \\
 & \phantom{Tab...10}$\blacktriangleright$ Image Generation & \cite{zhao2019variational}, \cite{mouches2021unifying}, \cite{zuo2021unsupervised}, \cite{zhao2023disentangling},
 \cite{ouyang2022disentangling} \\
 & \phantom{Tab...10}$\blacktriangleright$ Disease Classification & \cite{afshar2018brain}, \cite{wang2023deep} \\
  & \phantom{Tab...10}$\blacktriangleright$ Brain Age Regression & \cite{hu2020disentangled} \\
& \multicolumn{2}{l}{ Tab.~\ref{tab:hybrid_model_methods} \textbf{Hybrid models} } \\
 & \phantom{Tab...10}$\blacktriangleright$ Disease Classification & \cite{abuhmed2021robust}, \cite{lee2019toward}, \cite{liu2020masked}, \cite{nguyen2022interpretable}, \cite{qiu2020development}, 
 \cite{mohammadjafari2021using},
 \cite{mulyadi2023estimating}, 
 \cite{wolf2023don}, 
 \cite{kang2022prototype}, \cite{qiang2020deep} \\
 & \phantom{Tab...10}$\blacktriangleright$ Brain Age Regression & \cite{hesse2023prototype} \\
 & \phantom{Tab...10}$\blacktriangleright$ Clinical Score Regression & \cite{shimona2020deep} \\
& \multicolumn{2}{l}{ Tab.~\ref{tab:generative_methods} \textbf{Generative models} } \\
 & \phantom{Tab...10}$\blacktriangleright$ Disease Classification & \cite{baumgartner2018visual}, \cite{bass2020icam}, \cite{bass2022icam}, \cite{lanfredi2020interpretation}, \cite{liu2021going} \\
 & \phantom{Tab...10}$\blacktriangleright$ Brain Age Regression & \cite{bass2022icam} \\
 & \phantom{Tab...10}$\blacktriangleright$ Tissue Segmentation & \cite{bercea2023reversing},
 \cite{sanchez2022healthy}, \cite{wolleb2022diffusion} \\
& \multicolumn{2}{l}{ Tab.~\ref{tab:causal_model_methods} \textbf{Deep structural causal models} } \\
 & \phantom{Tab...10}$\blacktriangleright$ Image Generation & \cite{pawlowski2020deep}, \cite{reinhold2021structural}, \cite{rasal2022deep} \\
& \multicolumn{2}{l}{ Tab.~\ref{tab:attention_based_methods} \textbf{Attention-based models} } \\
 & \phantom{Tab...10}$\blacktriangleright$ Disease Classification & \cite{jin2020generalizable}, \cite{zhao2022attention}, \cite{sarraf2023ovitad} \\
 & \phantom{Tab...10}$\blacktriangleright$ Tissue Segmentation & \cite{gu2020net} \\
 & \phantom{Tab...10}$\blacktriangleright$ Brain Age Regression & \cite{dahan2022surface} \\
\end{tabular}
    \caption{Summary of both post-hoc and intrinsic methods}
    \label{tab:all_methods_v3}
\end{table}
\clearpage

\section{Methods}

\subsection{Post-hoc methods}\label{sec:posthoc_methods}

Post-hoc interpretability methods, as the name suggests, analyse model decisions after a network has been trained. While some post-hoc methods are model \textit{agnostic}, \textit{i.e.}, they can be applied to any \ac{ml} model, in some cases, they are only applicable to a specific family of models, such as \acp{cnn}. Agnostic post-hoc methods can be applied to ``black-box" models without requiring knowledge of the model parameters, as they generally analyse feature input and output pairs. Alternatively, post-hoc methods may require access to pre-trained model information (\textit{e.g.}, model weights) as for gradient-based and weight-analysis methods. \\

\subsubsection{Perturbation-based methods}\label{sec:perturbation_methods}

Perturbation-based methods explicitly alter the input features and measure the change in the model prediction between the original and perturbed data to discover relevant features. The most salient features for a model decision are those that produce the greatest change in the model prediction when perturbed. Perturbation-based methods mainly differ according to how they alter the input features.\\

\begin{figure}[h]
    \centering
    \includegraphics[width=.75\textwidth]{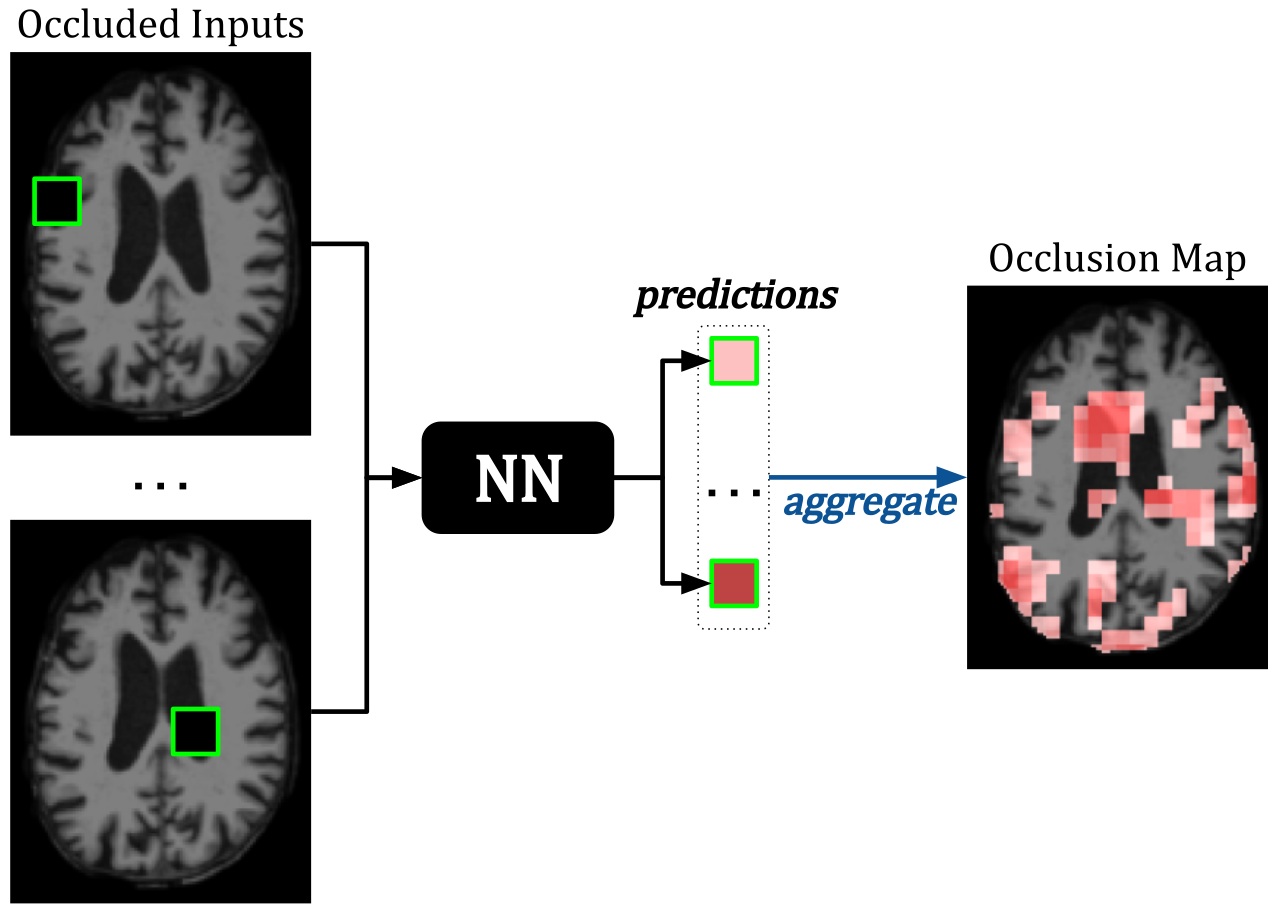}
    \caption{Example of\textbf{ \textit{Occlusion}} applied to an MRI image. In a patch-wise manner, a tile of the image is occluded, and the occluded image is fed to a neural network (NN) for prediction. The difference in predicted probability between the original and occluded image is assigned to the patch location in the occlusion map. Patches that result in the greatest change in prediction when occluded are interpreted as the most important for the model task \cite{zeiler2014visualizing}.}
    \label{Occlusion.png}
\end{figure}

Several perturbation-based methods occlude input features. For example, \textbf{\textit{Occlusion}} obstructs regions of an input image in a patch-wise fashion \cite{zeiler2014visualizing}. For every patch location, the change in the model output between the original and occluded image is calculated to form a sensitivity map. For classification tasks, sensitivity is the change in predicted probability $\mathbb{P}(c)$ of the image belonging to a class-of-interest $c$, as shown in Fig. \ref{Occlusion.png}. For regression tasks, the residual difference of the model prediction is assessed. \\

\textbf{\textit{Meaningful Perturbations}} follows a similar approach of occluding image regions but uses gradient descent to learn the occlusion mask that obfuscates the smallest region of the image that renders the model unable to correctly classify the masked image \cite{fong2017interpretable}. The masking process may replace pixel values with a constant value, Gaussian noise or by blurring. \\

Also incorporating occlusion, \textbf{\textit{Local Interpretable Model-Agnostic Explanations (LIME)}} approximates a ``black-box model" locally to an input $x$; then an interpretable \ac{ml} model, such as a linear model, is trained to mimic the ``black-box" model predictions for occluded samples of $x$ \cite{ribeiro2016should}. First, several perturbed images are generated from a given image $I_0$; a single perturbed image $I$ is generated by switching off a random subset of superpixels of $I_0$, where a superpixel is a set of neighbouring pixels with similar intensity. A sparse linear model is trained on the corresponding binary features $I' = (b_1, ..., b_n)$ where $b_i=0$ if superpixel is switched off to generate image $I$ and $b_i=1$ otherwise. Training labels for the linear model are the ``black-box" model predictions for perturbed images $I$. The feature importance of the $i^{th}$ superpixel in $I_0$ is given by the associated linear model coefficient of $b_i$. \\

In contrast to occluding image regions, several perturbation methods swap image regions or input features with those of another subject so that the altered image still appears realistic. Such an approach was proposed in the \textbf{\textit{Swap Test}}, where a reference image is selected that is from a different class to the image-of-interest \cite{nigri2020explainable}. For example, for an image classified as \ac{ad}, the reference image is randomly selected from healthy control images. In a patch-wise manner, a patch in the reference image is replaced with the corresponding patch in the image-of-interest and the change in model output between the reference and altered reference image is computed. The process is repeated for several randomly selected reference images and averaged. \\

Similarly, \textbf{\textit{Permutation Feature Importance}} \cite{fisher2019all} randomly permutes values of each input feature across samples. Let $P_\text{orig}$ be the model performance on the original data and $P_\text{perm}$ be the model performance when feature $j$ has been randomly permuted; then the importance of feature $j$ is either the ratio $P_\text{perm} / P_\text{orig}$ or difference $P_\text{perm} - P_\text{orig}$. The assumption is if feature $j$ is ignored by the model, then randomly shuffling feature $j$ will not influence model predictions. In contrast to previously mentioned perturbation methods, permutation feature importance is a global interpretability method. \\

\textbf{Advantages and disadvantages: }Perturbation-based methods have the advantage of being easy to implement and understand; they do not require a specific type of network nor access to the gradients. These methods may be applied to any ``black-box'' model, as they only need access to the input image and output value. However, these methods are computationally intensive and time-consuming, as inference is run for each location of the perturbation block. Another disadvantage is that perturbed images no longer belong to the training data distribution, so distribution shift may be responsible for any changes in model output rather than feature relevance \cite{hooker2019benchmark}. Concerning \textit{Occlusion}, this method is also sensitive to the size and the replacement intensity of the occluded patch \cite{fong2017interpretable}. 

\subsubsection{Gradient-based methods}\label{sec:gradient_methods}

Gradient-based methods compute the partial derivative of an output from a neural network output with respect to each input feature, using the backpropagation algorithm \cite{rumelhart1985learning}. The resulting gradient maps visualise how sensitive a neural network output is to small changes in input feature values, and they are also referred to as sensitivity maps. \\
\begin{figure}[h]
    \centering
    \includegraphics[width=.75\textwidth]{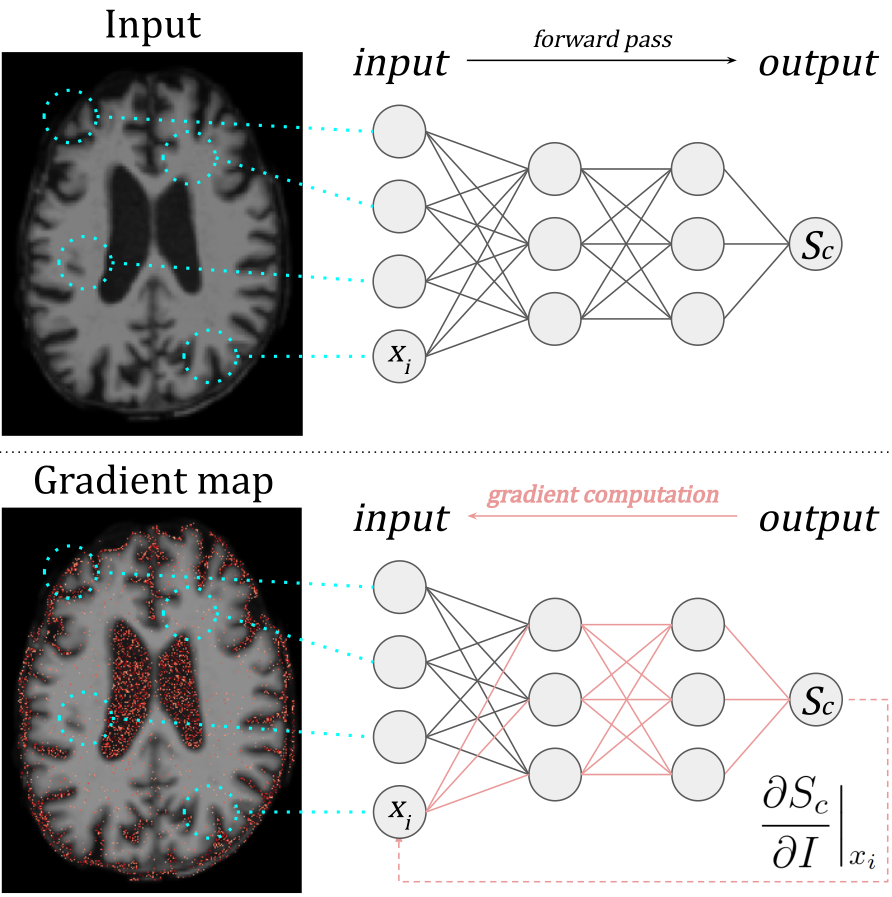}
    \caption{Example of \textbf{\textit{Vanilla Gradients}} applied to an MRI image. Partial derivatives for each voxel with respect to the network output score $S_c$ for class $c$ are computed. Pixels with the largest gradients are interpreted to have the greatest influence on the model prediction \cite{simonyan2013deep}.}
    \label{Gradient.png}
\end{figure}

\textbf{\textit{Vanilla Gradients}} was the first gradient-based method used to compute gradient maps for a \ac{cnn} trained to classify images \cite{simonyan2013deep} (see Fig. \ref{Gradient.png}). Let $I_0$ be an image with $N$ channels; $c$, a class-of-interest; and let $S_c(I)$ be the class score output function of a trained \ac{cnn} classifier. Then \textit{Vanilla Gradients} computes the absolute value of the partial derivative of $S_c(I)$ with respect to each voxel in $I_0$. Where $N>1$, the maximum value across channels is returned. \\

Two main limitations of \textit{Vanilla Gradients} exist: shattered gradients and the saturation problem. Firstly, gradient maps are often noisy because of ``shattered gradients'', where similar pixel values have substantially different partial derivatives of $S_c$, thus producing noisy maps \cite{balduzzi2017shattered}. Secondly, there is the ``saturation problem''. The function $S_c(I)$ learned by a \ac{cnn} is non-linear, therefore the \textit{Vanilla Gradient} map of $I_0$  does not interpret the behaviour of $S_c(I)$ globally, but locally to $I_0$. In particular, when $S_c(I)$ is saturated at $I_0$, \textit{i.e.} the gradient is close to zero, \textit{Vanilla Gradients} may not reveal image features that cause $S_c(I)$ to substantially change and switch predicted class \cite{shrikumar2017learning}. \\

\textbf{\textit{Grad $\times$ Input}} attempts to overcome the shattered gradients limitation through element-wise multiplication of \textit{Vanilla Gradients} with $I_0$, producing visually sharper sensitivity maps than \textit{Vanilla Gradients} \cite{kindermans1611investigating}. \\

\textbf{\textit{SmoothGrad}} was also developed to address the shattered gradients limitation of \textit{Vanilla Gradients} by adding random noise to the input image to create many noisy images, then computing the mean of the associated \textit{Vanilla Gradients} sensitivity maps \cite{smilkov2017smoothgrad}. \\
 
\textbf{\textit{Integrated Gradients}} addresses the saturation problem of \textit{Vanilla Gradients }\cite{sundararajan2017axiomatic}. Global behaviour is captured by travelling from a baseline image $I_b$  (\textit{e.g.}, an image of all zeros) to the image-of-interest $I_0$, and sampling $m$ images along the path: $I_b + \frac{k}{m} (I_0 - I_b) $ for all images $k$ from 1 to $m$. \textit{Integrated Gradients} then computes the mean \textit{Vanilla Gradients} map across the $m$ images (please see \ref{appendix:A}). Notably, \textit{Integrated Gradients} tends to highlight more relevant image features compared to \textit{Vanilla Gradients} and \textit{SmoothGrad}. However, \textit{Integrated Gradients} maps may still include noisy gradients from saturated regions of $S_c(I)$ \cite{miglani2020investigating}. \\

\textbf{Advantages and disadvantages: }Gradient-based methods are fast to run and easy to understand. However, in addition to the shattered gradients and saturation problem previously discussed, gradient maps are less able to discriminate between classes than other interpretable methods.
    
\subsubsection{Backpropagation-based methods}\label{sec:backpropagationbased_methods}

Backpropagation-based methods apply rules other than gradients to map the output score back to the input features to assign feature relevance. The earliest backpropagation methods for \acp{cnn} were identical to \textit{Vanilla Gradients} aside from their treatment of the \ac{relu} function. 

Specifically, \textit{Vanilla Gradients} back-propagates through a \ac{relu} function by setting a gradient value to zero if the corresponding value in the forward feature map is negative. In comparison, \textit{\textbf{\textit{Guided Backpropagation}} \cite{springenberg2014striving}} performs the same operation and also sets negative gradients to zero. Consequently, \textit{Guided Backpropagation} only allows positive gradients, whereas \textit{Vanilla Gradients} may produce negative gradients. \\


\begin{figure}[h]
    \centering
    \includegraphics[width=.75\textwidth]{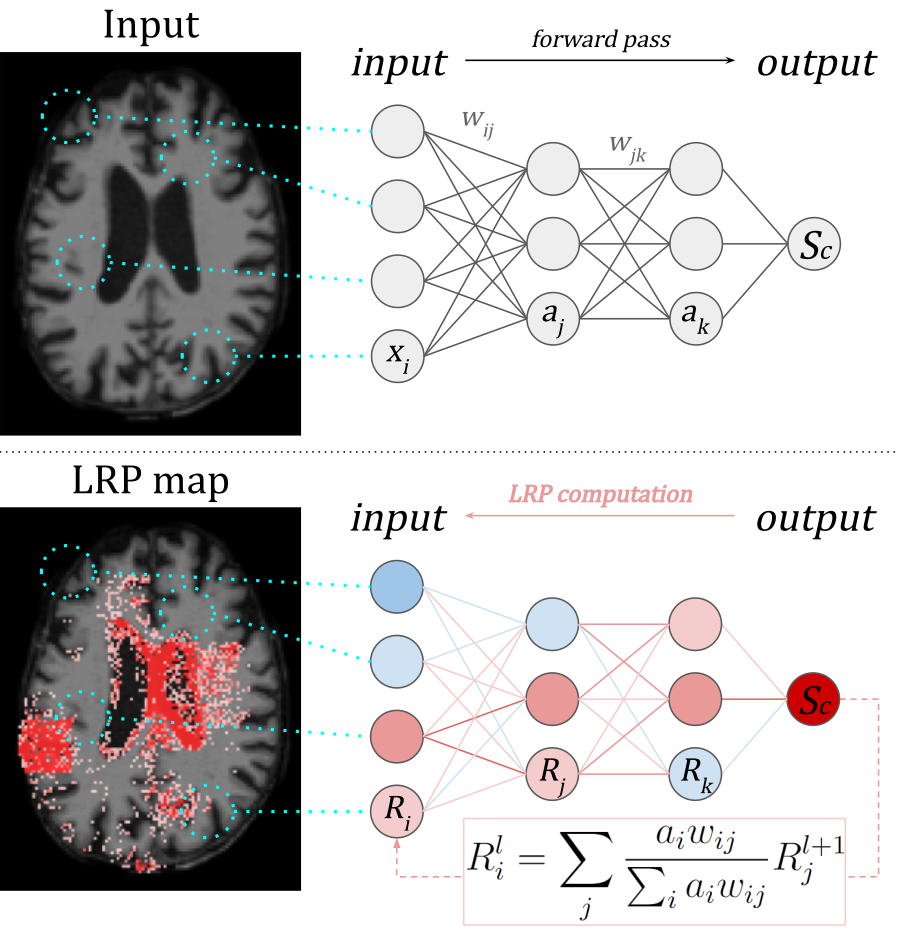}
    \caption{Example \textbf{\textit{Layer-wise Relevance Propagation} (LRP)} applied to an MRI image. The network output score $S_c$ for class $c$ is redistributed backwards through the network according to the equation shown until the input image is reached. The pixels with the highest proportion of $S_c$ are interpreted as having the greatest contribution to the model prediction \cite{bach2015pixel}.}
    \label{LRP.png}
\end{figure}



\textbf{\textit{\Ac{lrp}}} is another popular backpropagation method \cite{bach2015pixel}, as visualised in Fig. \ref{LRP.png}. In \textit{\ac{lrp}}, the model output score $S_c(I_0)$ is redistributed backwards through the network, layer by layer, until the input image $I_0$ is reached. Each node (or pixel) is allocated a relevance value, which is the weighted sum of relevance values of connected nodes in the neighbouring higher layer. Different \ac{lrp} rules have a different weighted sum based on the network parameters, but all follow the relevance conservation principle: relevance assigned to a node from the neighbouring higher layer is equal to the relevance passed from that node to the neighbouring earlier layer (for more information, please see \ref{appendix:B}) \\

\textbf{Advantages and disadvantages:} Analysis carried out by Adebayo \etal{} demonstrated \textit{Guided Backpropagation} maps are independent of higher network layer parameters and sample labels, which is undesirable for an interpretability method \cite{adebayo2018sanity}. Additionally, \textit{\ac{lrp}} is sensitive to hyperparameter selection and may be difficult to tune.

\subsubsection{Class activation maps}\label{sec:cam_methods}

\begin{figure}[h]
    \centering
    \includegraphics[width=.75\textwidth]{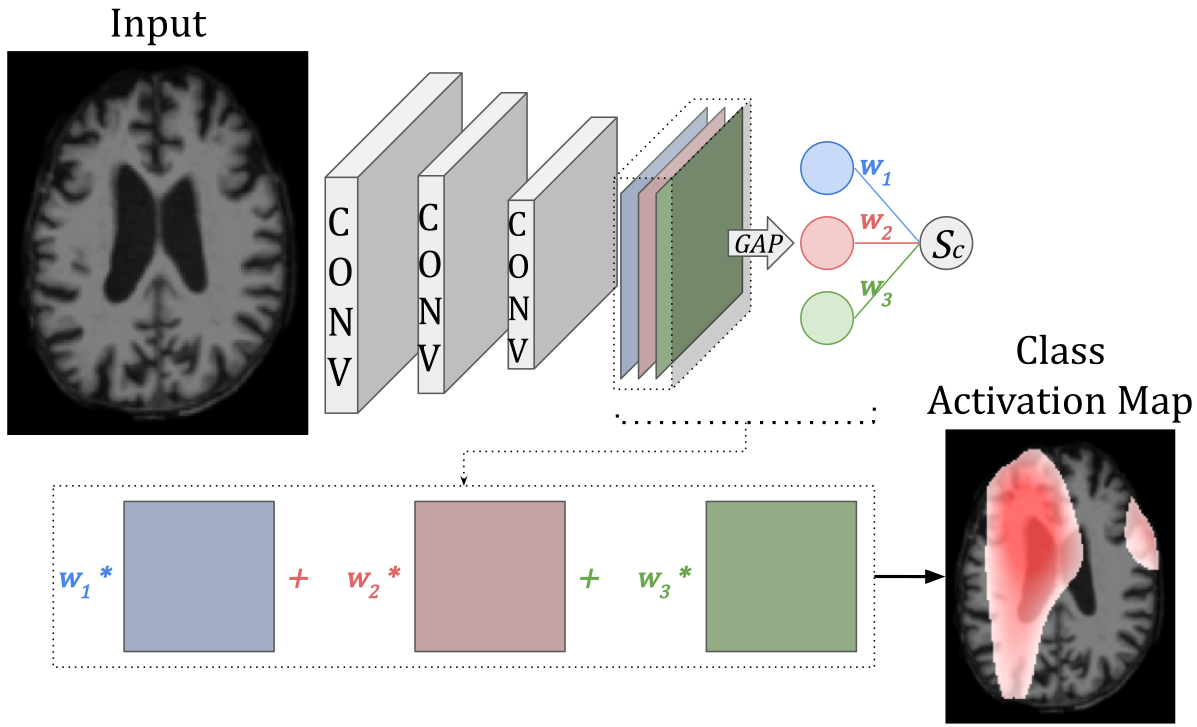}
    \caption{Example of a \textbf{Class Activation Map (CAM)}-based method where the activation maps of the final convolutional layer are weighted by the weights computed through the global average pooling (GAP) layer to produce a class activation map. Image adapted from \myetal{Zhou} \cite{zhou2016learning}.}
    \label{CAM.png}
\end{figure}

\textbf{\textit{Class Activation Maps (CAM)}} highlight image regions used by the final layer of a \ac{cnn} to classify the input image \cite{zhou2016learning}. To compute \textit{CAM} visualizations, the final layer of the network is required to be a \ac{gap} layer. In a \ac{gap}-\ac{cnn}, the weighted sum of the activation maps in the final layer determines the class score $S_c$ for each class $c$ (Eq.~\ref{eq:cam1}):

\begin{equation}\label{eq:cam1}
S_{c} =\sum_{k} w_{k}^{c} \sum_{x, y} A^{k}_{x, y} =\sum_{x, y} \sum_{k} w_{k}^{c} A^{k}_{x, y}
\end{equation}

where $A^{k}_{x,y}$ represents the activation of node $k$ in the last convolutional layer of the network at pixel location $(x,y)$, and $w^{c}_{k}$ represents the importance of node $k$ for the classification of class $c$ (see Fig. \ref{CAM.png}).\\

Then \textit{CAM} for class $c$ is defined as (Eq.~\ref{eq:cam2}):

\begin{equation}\label{eq:cam2}
CAM^{c}=\sum_{k} w_{k}^{c} A^{k}_{x, y}
\end{equation}

Hence, the sum of all elements in $CAM^{c}$ is equal to the class score $S_{c}$. \\

\textbf{\textit{Gradient-Weighted Class Activation Maps (Grad-CAM)}} extends \textit{CAM} to all \ac{cnn}s to obviate the need for a \ac{gap} layer \cite{selvaraju2017grad}. In \textit{Grad-CAM}, the weight $w_k^c$ is not learned as in a \textit{GAP-CNN}, but computed as the mean gradient of the score class $S_{c}$ with respect to activation map $A^{k}_{x,y}$ of a layer-of-interest (usually the last layer). Then \textit{Grad-CAM} visualises features with positive influence only (Eq.~\ref{eq:cam4}):

\begin{equation}\label{eq:cam4}
{\text{Grad-}CAM}^{c}={Re}{L} U\left(\sum_{k} w_{k}^{c} A_{x, y}^{k}\right)
\end{equation}

Finally, the \textit{CAM} or \textit{Grad-CAM} heatmap is up-sampled to the original input image size and superimposed on the input image, which is why these heatmaps have a coarse resolution. \\


\textbf{Advantages and disadvantages: }\textit{Grad-CAM} is a popular method of interpretability, both for natural images and medical images. It is most often applied to image classification since the heatmaps are class-specific, but it can also be applied to regression and segmentation tasks. \textit{Grad-CAM} does not require a modified \ac{cnn} architecture; is not computationally intensive; is easy to implement and widely available in multiple libraries. A disadvantage of \textit{CAM} and \textit{Grad-CAM} is that the heatmaps are coarse (low resolution) because they are often upsampled from the last convolutional layer of a network. To improve the resolution, \textit{Grad-CAM} has been coupled with other pixel-wise attribution methods such as \textit{Guided Backpropagation}, known as \textbf{\textit{Guided Grad-CAM}}. In \textit{Guided Grad-CAM}, the \textit{Grad-CAM} output is multiplied element-wise with the \textit{Guided Backpropagation} heatmap. \\

\subsubsection{Weight analysis}
An alternative approach for visualizing and explaining the decisions of a network is to analyse the weights of the trained network. However, as deep neural networks learn high-level features in the hidden layers, simply visualizing the raw learned features usually does not offer human-interpretable explanations \cite{molnar2020interpretable}.
Weight analysis methods attempt to create human-understandable explanations through clustering weights and associating clusters with human concepts. \\

\begin{figure}[h]
    \centering
    \includegraphics[width=.75\textwidth]{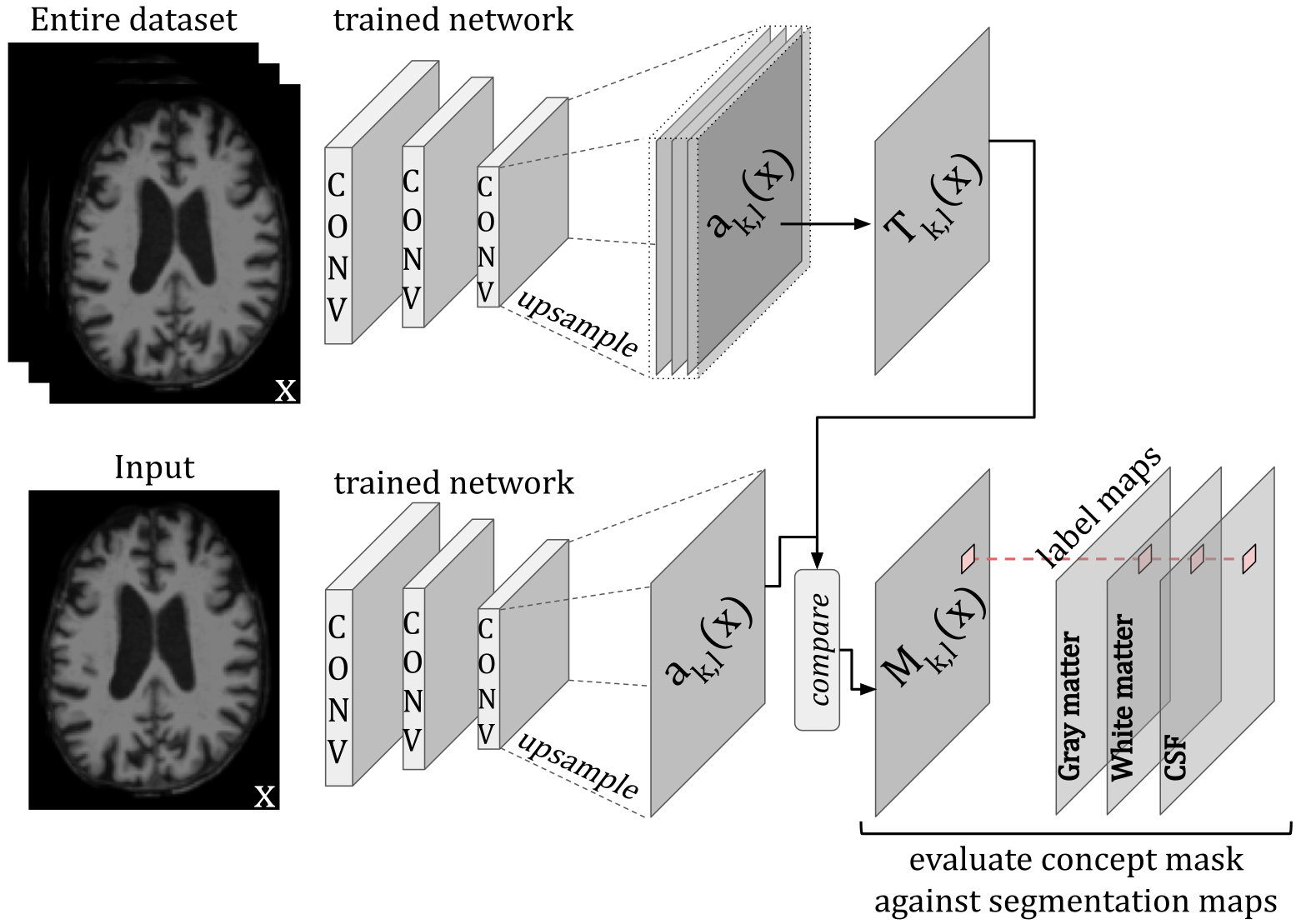}
    \caption{Example of a \textbf{Network Dissection} model where the activation map of individual filters in the network are analysed to identify which specific concepts they have learnt by evaluating them against segmentation maps. Image adapted from \myetal{Bau} \cite{bau2017network}.}
    \label{Weight.png}
\end{figure}

The \textbf{\textit{Network Dissection}} approach \cite{zhou2019comparing} quantifies the interpretability of a \ac{cnn} by evaluating the alignment between activated regions of individual hidden filters and human-labelled concepts (objects, parts, textures, colours). The process involves first defining a set of task-relevant concepts and then creating annotation masks $L_{c}(x)$ for each concept $c$ and image $x$. Next, masks $M_{k}(x)$ of the top activated areas per filter $k$ and per image $x$ are created by scaling the activation maps $A_{k}$ to the size of the input images, and binarizing them (thresholding on the top quantile level $T_{k}$ of the distribution of pixel activations for filter $k$ over all images). Finally, the accuracy of each filter $k$ in detecting concept $c$ is reported as the sum of the Intersection over Union (IoU) between $M_{k}(x)$ and $L_{c}(x)$ across all the images in a dataset (see Fig. \ref{Weight.png}). To quantify the interpretability of a layer, the number of unique concepts aligned with filters, \textit{i.e.}, unique detectors, are counted.  \\


A graphical representation of the concepts learned by a network to understand its behaviour was proposed \cite{kori2020abstracting}. This \textit{\textbf{Concept Graphs}} framework involves grouping similar weight vectors through hierarchical clustering in order to define concepts. Then, formed weight clusters are associated with some region in the input image by using a variation of \textit{Grad-CAM}; the region corresponds to a human-understandable concept, for example, a tumour boundary. After the concepts have been identified, a concept graph is formed that represents the link between concepts in different layers. This is computed by intervening on the pairs of concepts and calculating the mutual information (MI) between pre-interventional and post-interventional distributions as a measure of the link between two concepts. The trails of concepts on the graph, therefore, represent the flow of information used by the network when making a prediction. \\

Several papers integrate \textbf{\textit{Community Detection}} within a \ac{dl} model for analysing \ac{fmri} data \cite{dvornek2019jointly, li2021braingnn}. The aim of \textit{Community Detection} in the context of neuroimaging is to discover $K$ networks of brain regions that are salient for a particular \ac{dl} task. Given an \ac{fmri} connectivity matrix defined over $N$ brain regions, the \ac{dl} model incorporates a fully connected layer with a weight matrix $W \in \mathbb{R}^{N \times K}$. Each value $w_{nk} \in W$ may be interpreted as a membership score of brain region $n$ belonging to the community $k$. A clustering algorithm is then applied to the weights to assign brain regions to communities. \\

\subsection{Intrinsic methods}\label{sec:intrinsic_methods}

Intrinsic interpretability refers to \ac{ml} models that are explainable by design, \textit{i.e.}, where feature representations can be understood by humans. The interpretability can be due to the simple structure of the models, such as short decision trees or sparse linear models, where network decisions can be easily followed. Alternatively, interpretability can be achieved by explicitly including interpretable modules or constraints in the model, as is required for designing \ac{idl} models. In this section, we present five categories of intrinsic interpretable methods: disentangled latent spaces, interpretable hybrid models and interpretable intermediate features, interpretable generative models, deep structural causal models, and attention mechanisms.  \\

\subsubsection{Disentangled latent spaces}\label{sec:intrinsic_disentangled}

\begin{figure}[h]
    \centering
    \includegraphics[width=.75\textwidth]{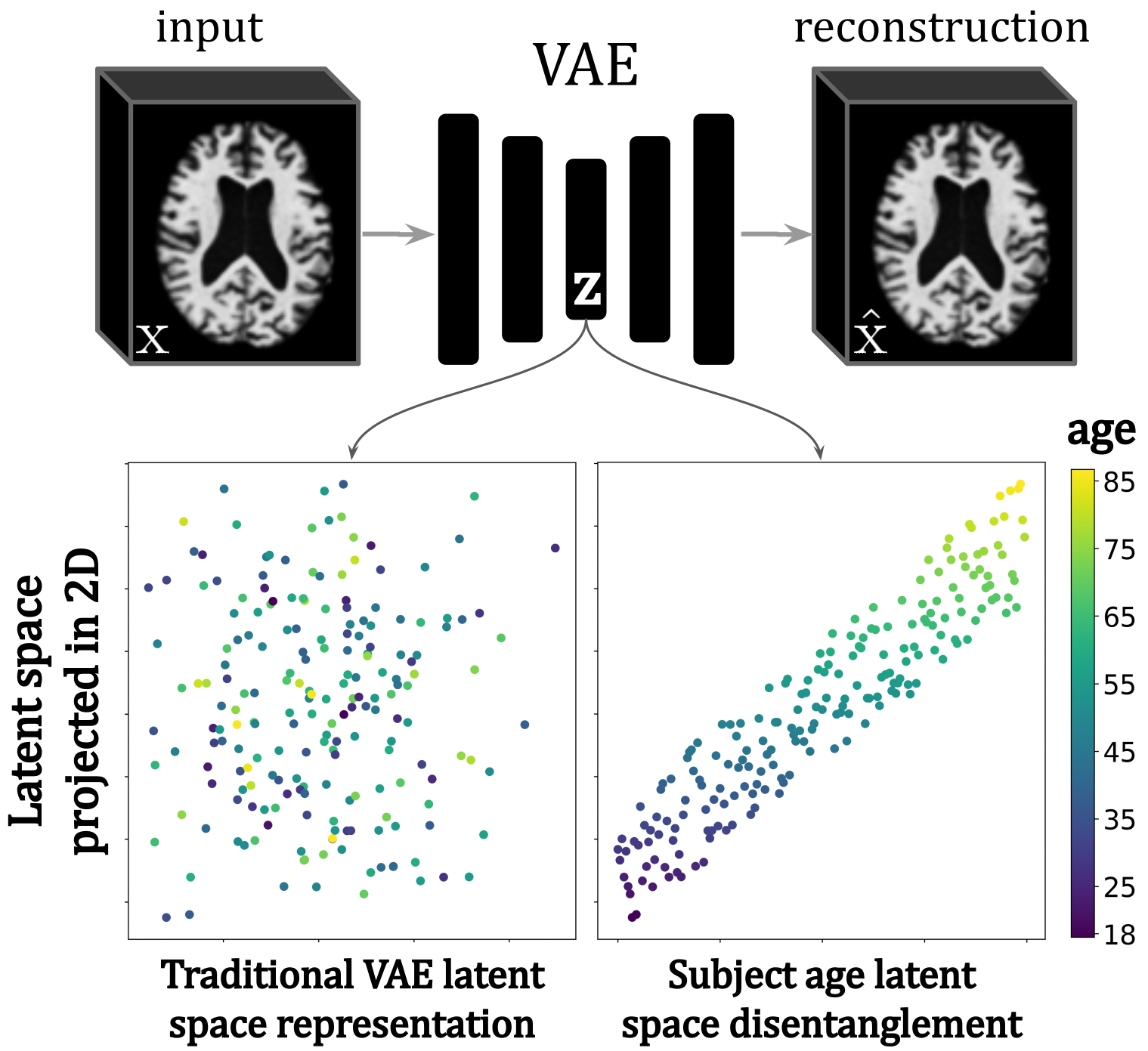}
    \caption{Example of a traditional \textit{vs.} a \textbf{disentangled latent space} of a trained \ac{vae} where age was added as a condition. The structured latent space can, therefore, be used to generate new samples for a given condition (such as age), as well as understand what type of changes occur in a given brain image with age. Image adapted from \myetal{Zhao} \cite{zhao2019variational}. }
    \label{Latent.png}
\end{figure}

The latent space of a neural network is a learned representation of the input data that has usually undergone compression, such that similar input samples are transformed into representations that are close together in this space. A popular \ac{dl} model is the \ac{ae}, where an encoder learns to compress input data to a latent space, and a decoder learns to reconstruct the input data from the latent representations. An extension to the \ac{ae} that enables data generation from the latent space is the \ac{vae}, where the latent space is constrained to a multivariate Gaussian distribution \cite{kingma2013auto}. A desirable property is that the latent space is to some extent \textit{disentangled}, meaning a single factor in the latent space corresponds to a single source of variation in the high-dimensional image space. This can be encouraged through the introduction of losses, which optimise for a subset of the latent space to encode specific semantic feature(s) in the image space. This is illustrated in Fig~\ref{Latent.png}, showing a traditional \textit{vs.} disentangled (for subject age) latent space of a trained \ac{vae}. Note that the structure of the latent space projection in 2D space is more coherent for the disentangled space than it is for the traditional space. \\


\textbf{\textit{Capsule Networks}} are an alternative architecture to \acp{cnn} that learn disentangled, interpretable activation vectors \cite{sabour2017dynamic}. \textit{Capsule Networks} learn spatial relationships between an object and its constituent parts, which are invariant to the object viewpoint. Elements of an activation vector learn pose parameters for an associated object, such as size, orientation, texture, and hue. The $L_2$-norm (equal to the Euclidean distance from the origin) of an activation vector is equal to the predicted probability of the corresponding object, thus enabling classification. \\

\textbf{Advantages and disadvantages: }Disentangled latent representations provide some control for image generation to the end user. The user can manipulate features in the generated image in a semantically meaningful way by interpolating a disentangled factor in the latent space. One limitation of disentangling latent spaces for complex data is that the generative factors may not be inherently independent, and by constraining the latent representation to have independent representations, useful information about these dependencies can be lost \cite{mathieu2019disentangling}. Additionally, constraining the latent space representations often comes at the expense of performance \cite{higgins2017betavae}. One disadvantage of ProtoPNet, in particular, is that distance maps are upsampled from the latent space to the image space, which implicitly assumes that spatial relationships in the image space are preserved in the latent space. However, Wolf \etal{} \cite{wolf2024keep} proved this is not necessarily the case, though efforts are being made to account for this issue \cite{wolf2024keep, carmichael2024pixel}. \\

\subsubsection{Interpretable hybrid models and interpretable intermediate features}\label{sec:intrinsic_hybrid}

A \textit{hybrid} \ac{dl} model usually has two components: a \ac{nn} that learns intermediate feature representations from the input data, coupled with a model that predicts the learning task from the feature representations. The second component can either be a \ac{nn} or some other \ac{ml} model, referred to as \ac{nn} + \ac{nn} and \ac{nn} + \ac{ml} hybrid model, respectively. \\

\begin{figure}[!h]
    \centering
    \includegraphics[width=1\textwidth]{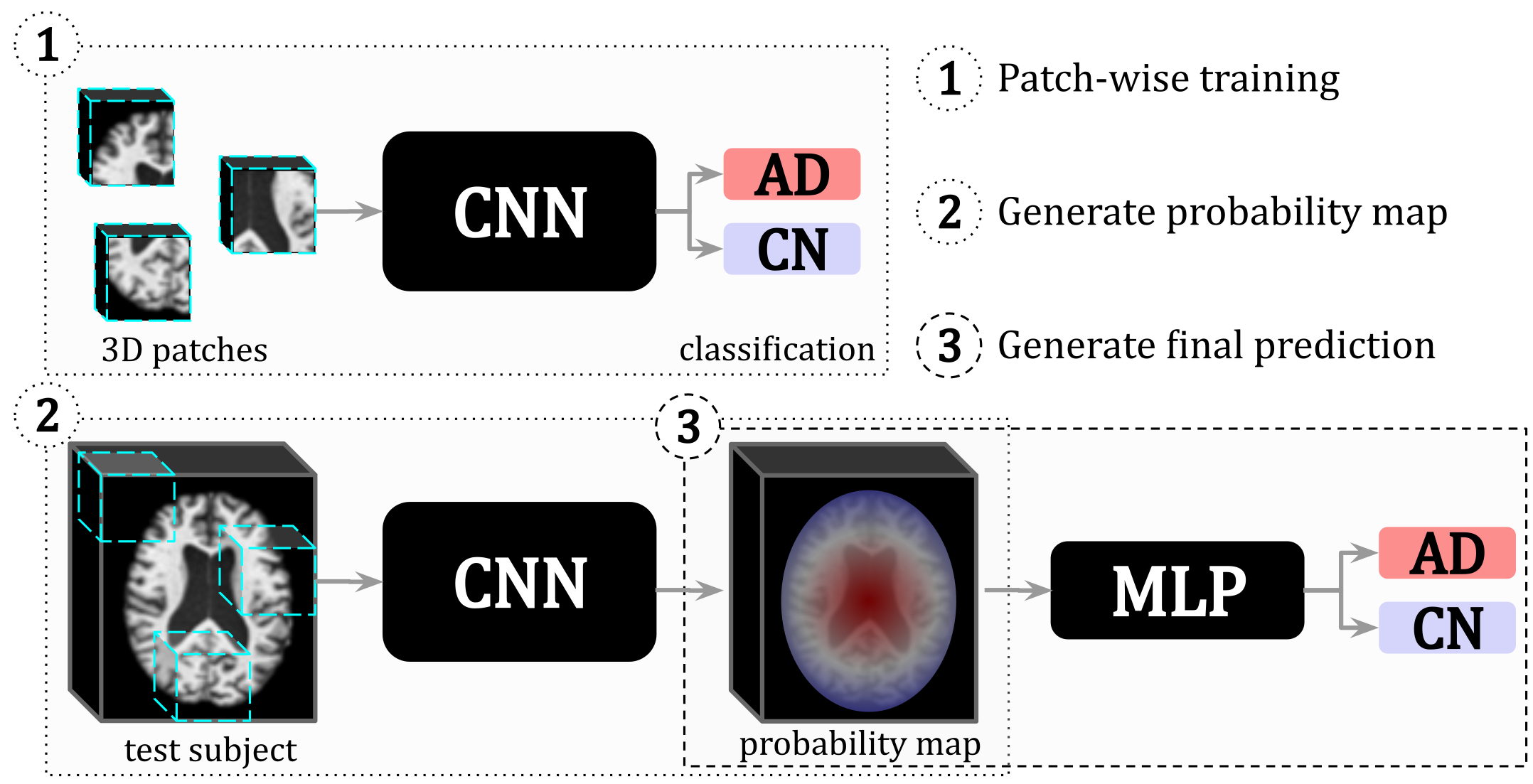}
    \caption{Example of an \textbf{interpretable hybrid model} where the intermediate probability map is used as the features for an \ac{mlp} model and acts as the model explanation.
    The authors of this study \cite{qiu2020development} proposed a three-step approach: first, a \ac{cnn} classifier was trained to predict whether a given 3D brain \ac{mri} patch is \ac{ad} or \ac{cn} (1), then, the trained \ac{cnn} produced a probability map (the intermediate feature) for a given test subject (2), and, finally, an \ac{mlp} was trained on the intermediate feature probability map to distinguish between \ac{ad} and \ac{cn} (3).
    Image adapted from \myetal{Qiu} \cite{qiu2020development}.}
    \label{fig:hybrid}
\end{figure}

An \textit{interpretable hybrid model} is a hybrid model that possesses intermediate feature representations that can be understood by humans and, therefore, act as model explanations (see Figure~\ref{fig:hybrid}). Some researchers also compute the feature importance of the second model component and thus generate a second set of model explanations along with the intermediate features \cite{abuhmed2021robust, lee2019toward}. If a study computes feature importance as a second set of explanations, we refer to their approach as ``int. features + feature importance". \\


One notable example of an interpretable hybrid model is the prototypical part network (ProtoPNet), which mimics human reasoning when classifying an image \cite{chen2019looks}. The network learns a fixed number of \textit{prototypes} for each class, where a \textit{prototype} is a tensor in latent space that is associated with an image patch containing features typical of that class. At test time, latent features of an image are compared to each \textit{prototype} by computing a maximum similarity score, and the similarity scores are passed through a fully connected layer to predict the image class. Several studies in our review employed \textit{prototype} layers in their model architecture, inspired by ProtoPNet \cite{mohammadjafari2021using, wolf2023don, mulyadi2023estimating}. \\


\textbf{Advantages and disadvantages: }An advantage of interpretable hybrid models is that they may be designed so the intermediate features are suited for a particular application. For example, in a clinical setting, intermediate diagnostic features may be learned that are familiar to clinicians. However, interpretable hybrid models require careful design and may take a long time to develop. \\

\subsubsection{Interpretable generative models}\label{sec:intrinsic_generative}

\begin{figure}[h]
    \centering
    \includegraphics[width=.65\textwidth]{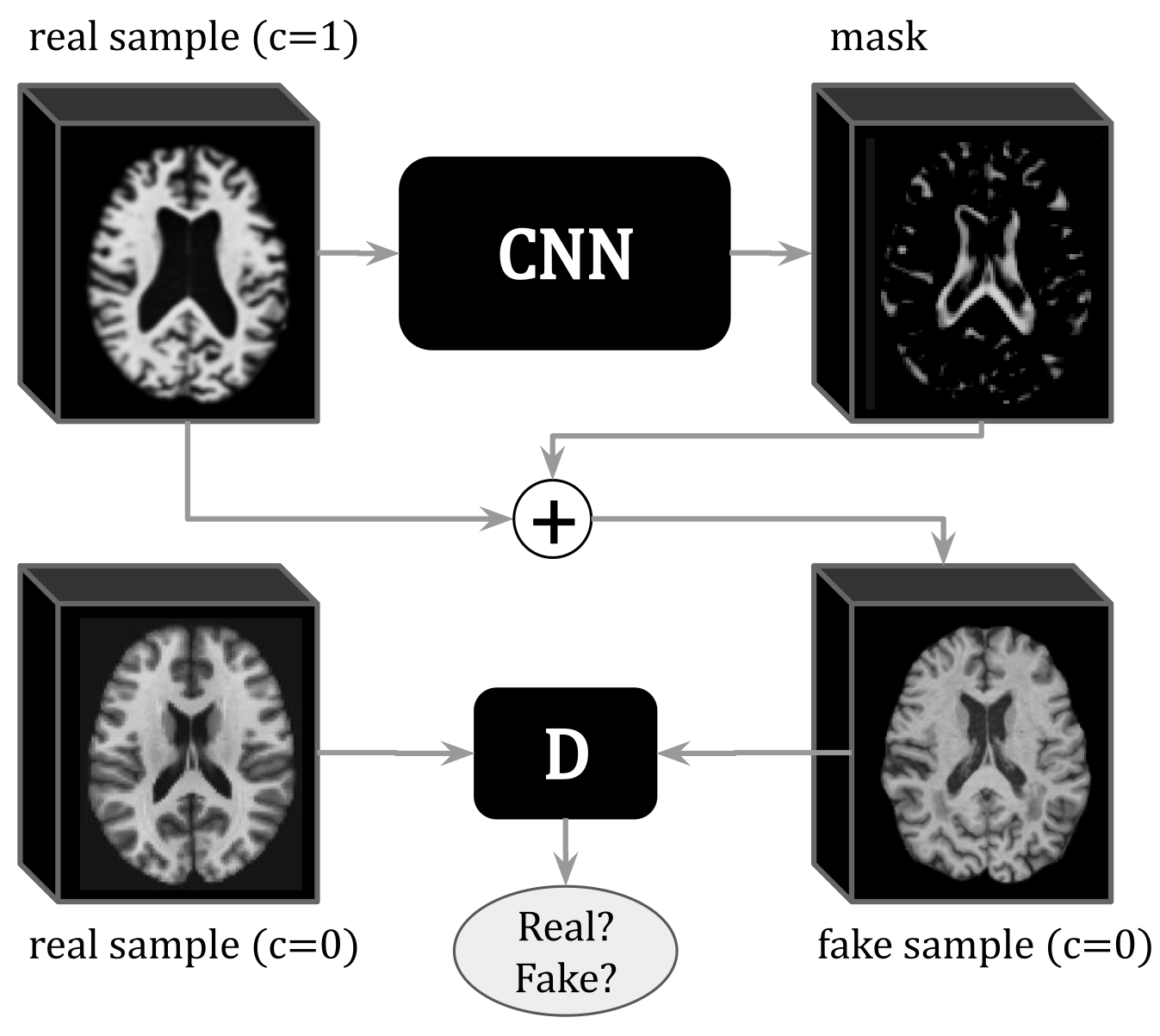}
    \caption{Example of a \textbf{ interpretable generative model} where a generator network (shown as \textbf{``CNN"} in the figure) produces the minimum mask needed to change the class of the input sample (from $c = 1$ to $c = 0$). 
    A discriminator network (\textbf{``D"}) is trained to distinguish between fake and real samples of the same class c in order to constrain the generator to produce realistic samples.
    Moreover, the masks can be used as explanations for the class discriminative features. Image adapted from \myetal{Baumgartner} \cite{baumgartner2018visual}. }
    \label{Generative.png}
\end{figure}

Another interpretability approach is to train a generative model to generate explanations for neuroimaging tasks. The model learns to generate modifications to the input image so that the modified image appears to belong to a different class. The modifications are then used as explanations for the prediction task. For example, in binary classification, the model $f$ modifies an input image $x$ of class $0$, such that $x' = f(x)$ appears to be from class $1$. This task is often referred to as anomaly detection or counterfactual generation, and the modifications are known as the anomaly map or disease effect map. Such an example is shown in Fig.~\ref{Generative.png} where a network produces the minimal additive mask needed to change an image from one class, \ac{ad} in this case, to another \textit{e.g.}, \ac{cn}. \\



\textbf{Advantages and disadvantages: }By learning to generate new images as explanations for discriminative tasks, generative methods are capable of capturing more meaningful class-discriminative features in comparison to methods that evaluate the features learned by classification networks. These generative methods also provide a framework to investigate how changing features in an image \textit{e.g.}, by interpolation, affects the network decision. However, generative models can be challenging to train and require high computational power, rendering these methods harder to implement.


\subsubsection{Deep structural causal models}\label{sec:intrinsic_causal}

\begin{figure}
    \centering
    \includegraphics[width=14cm]{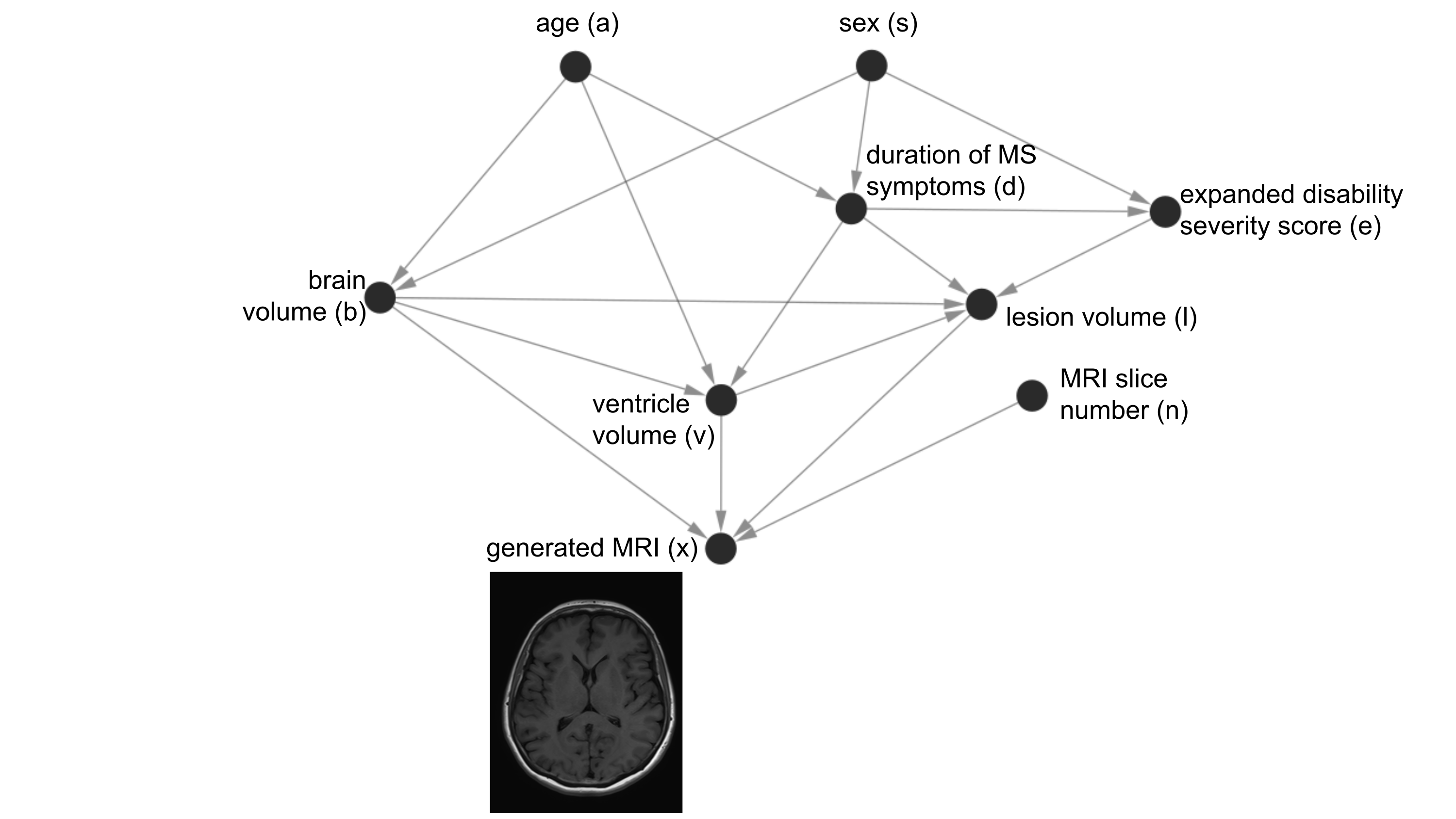}
    \caption{\textbf{Example of a causal graph where assumptions about the image generating mechanism are explicitly defined.} Deep structural causal models can then be learned to estimate \ac{mri} images under counterfactual scenarios.  Image adapted from Reinhold \etal{} \cite{reinhold2021structural} .}
    \label{fig: causal}
\end{figure}

Where randomised controlled trials are impossible, infeasible, or unethical, estimating causal effects is often still possible using causal inference methods. One such method is the \textit{\ac{scm}}, which estimates causal effects by simulating population-level interventions \cite{pearl2009causality}. An \ac{scm} consists of a set of $d$ endogenous variables $\{X_{1},..., X_{d}\}$, exogenous or noise variables $\{N_{1},..., N_{d}\}$, and structural assignments (denoted as $:=$):  

\begin{equation} 
X_{j} := f_{j}(\boldsymbol{PA_{j}}, N_j), j = 1,...,d 
\end{equation} 

where $\boldsymbol{PA_{j}} \subseteq \{X_{1},..., X_{d}\} \setminus \{X_j\}$ are the parents of $X_{j}$. The joint probability distribution over the noise variables is assumed to be jointly independent. An \ac{scm} has an associated causal graph $\mathcal{G}$ that visually represents our assumptions regarding how data were generated in the real world. The causal graph $\mathcal{G}$ is a directed acyclic graph (DAG) where all endogenous variables are represented as nodes. A directed edge $X_i$ $\rightarrow$ $X_j$ exists in $\mathcal{G}$ if $X_j$ depends on $X_i$ for its value. Indeed, we define $X_i$ to be a direct cause of $X_j$ if $X_i$ appears in the structural assignment $f_j$ for $X_j$. Fig.~\ref{fig: causal} is an illustration of a causal graph for \ac{ms}.\\ 

For a \ac{scm}, the causal effect of intervening on a variable $X$ by setting it to $a$ is denoted $do(X = a)$. It is also possible to estimate \textit{counterfactual} scenarios for specific individuals, which are hypothetical alternative outcomes to the actual outcome. We refer readers to \cite{pearl2016causal} and \cite{peters2017elements} for a detailed overview of \acp{scm}. \\

\textit{\acp{dscm}} employ neural networks to learn at least one of the structural assignments in the \ac{scm}, and applying them to medical imaging data is an emerging research topic \cite{castro2020causality, pawlowski2020deep, reinhold2021structural, rasal2022deep}. Pawlowski \etal{} trained a \textit{\ac{dscm}} on UK Biobank data to understand how a subject's age ($a$), sex ($s$), brain volume ($b$), and ventricle volume ($v$) influenced their brain \ac{mri} image ($img$) \cite{pawlowski2020deep}. The structural assignments were defined as (Eq.~\ref{eq:pawlowski_dscm}):

\begin{equation}\label{eq:pawlowski_dscm}
    \begin{split}
        b &:= f_1(a, s, N_1) \\
	v &:= f_2(a, b, N_2) \\
        img &:= f_3(v, b, N_3)      
    \end{split}
\end{equation}

where $N_1, N_2, N_3$ are noise variables. In this study, $f_1$ and $f_2$ were modelled with normalising flows, and $f_3$ was learned using a conditional \ac{vae} where the \ac{vae} generated an estimated brain MRI. A normalising flow is a sequence of invertible transformations $g = f_1 \circ f_2 \circ \text{...} \circ f_K$ that transforms a tractable distribution $z$ into a more complex distribution $x = g(z)$ (we refer readers to \cite{kobyzev2020normalizing} for an introduction to normalising flows). Images were generated for a variety of counterfactual scenarios, and difference maps between the generated and original images were visually inspected for interpretation. For instance, for a 49-year-old subject, an image was generated for the counterfactual $do (\text{age} = 80\text{year-old})$; the generated image exhibited increased ventricle volume and reduced brain volume compared to the original image, consistent with trends in the true distribution. \\

\textbf{Advantages and disadvantages: }The strength of \textit{\acp{dscm}} is that causal mechanisms of imaging markers may control for confounders, unlike most other \ac{dl} models. However, the causal graph $\mathcal{G}$ must be carefully constructed from domain knowledge, and the structure of $\mathcal{G}$ may not yet be fully elucidated. Furthermore, it is impossible to obtain ground truth data for counterfactual scenarios, so counterfactual images cannot be validated. \\

\subsubsection{Attention mechanisms}\label{sec:intrinsic_attention}
\begin{figure}[h!]
    \centering
    \includegraphics[width=.95\textwidth,keepaspectratio]{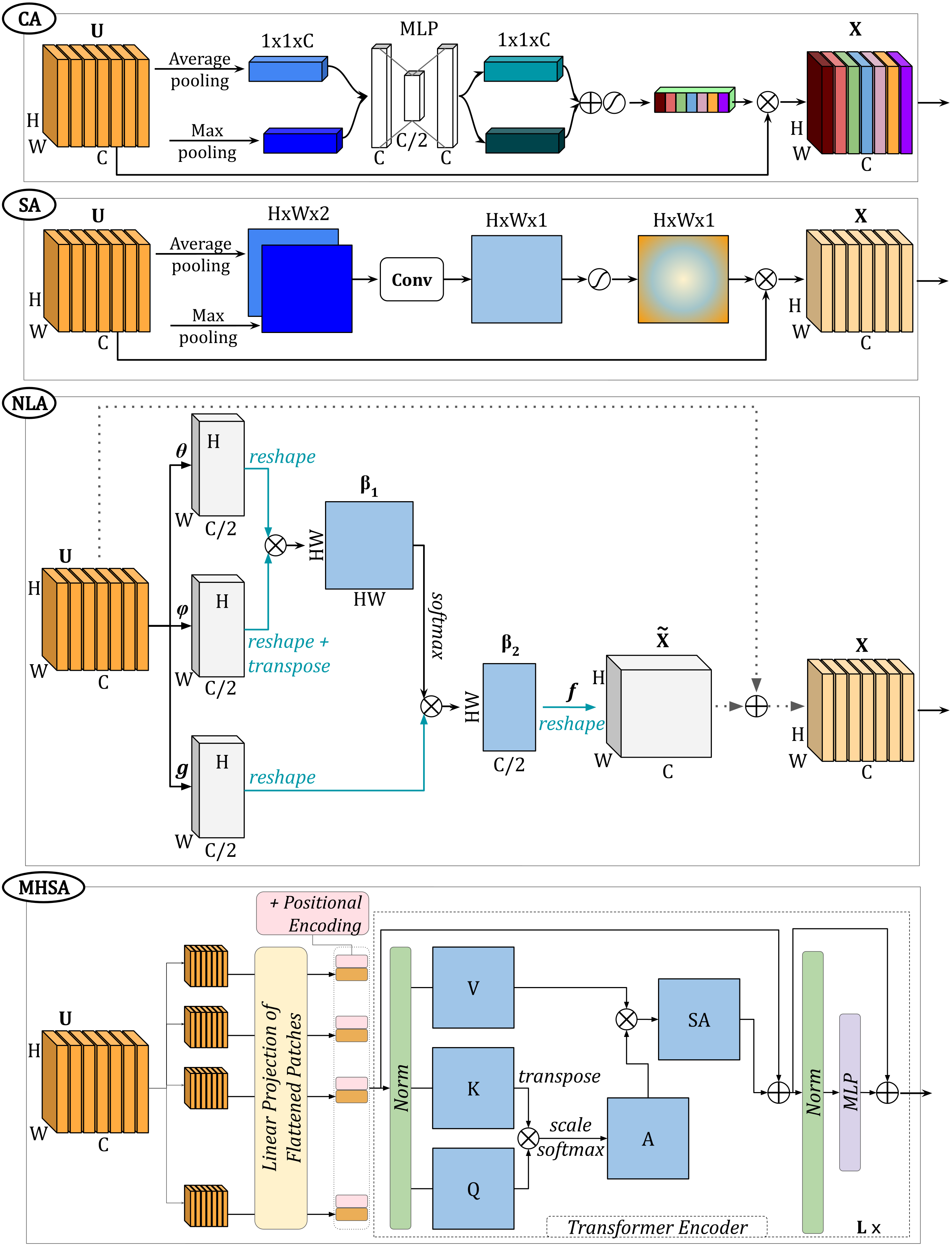}
    \caption{Example of \textbf{attention–based methods} showing channel attention in the first panel (\textbf{``CA"}), spatial attention in the second panel (\textbf{``SA"}), non-local attention in the third panel (\textbf{``NLA"}), and multi-head self-attention in the fourth panel (\textbf{``MHSA"}). 
    Note that \textbf{``CA"} and \textbf{``SA"} are adapted from \myetal{Woo} \cite{woo2018cbam}, \textbf{``NLA"} is adapted from \myetal{Wang} \cite{wang2018non}, while the fourth panel is adapted from \myetal{Dosovitskiy} \cite{dosovitskiy2020image}.}
    \label{Attention.png}
\end{figure}

In recent years, attention in the context of deep learning has become an important area of research as it can be easily incorporated into existing neural network architectures whilst also improving performance \cite{brauwers2021general,niu2021review} and providing explanations \cite{singh2020explainable, samek2017explainable}. \textbf{\textit{Attention}} methods learn a heatmap over the inputs, features, or channels of the neural network, subsequently used to weight the data to emphasise key features. In the following, we discuss four main types of \textit{attention}: channel, spatial, non-local, and self-attention, which are illustrated in Fig.~\ref{Attention.png}. For a more comprehensive description of \ac{dl} \textit{attention} mechanisms, we refer the reader to \myetal{Niu} \cite{niu2021review} and \myetal{Guo} \cite{guo2022attention}. \\
 
\textbf{\textit{Channel attention}} assigns a weight to each filter in order to emphasize useful features. One of the most popular \textit{channel attention} blocks is the \textit{squeeze-and-excitation} block \cite{hu2018squeeze}. Let $U$ be a feature map with dimensions $H \times W \times C$, then the \textit{squeeze-and-excitation} block comprises a \textit{squeeze} function ($F_{sq}$), which performs global average pooling \cite{lin2013network}, followed by an \textit{excitation} function ($F_{ex}$) defined as the sigmoid function ($\sigma$) applied to an \ac{mlp}. More specifically, the \textit{squeeze-and-excitation channel attention} $\alpha_{\mathbf{SE}}$ is defined as (Eq.~\ref{eq:attention_channel1}):

\begin{equation}\label{eq:attention_channel1}
    \begin{split}
    \alpha_{\mathbf{SE}} = \, & F_{ex} ( \, F_{sq} (U) \, )\\
    = \, & \sigma \left( W_2 \, \text{ReLU} \left( W_1 \, \, \text{GlobalAvgPool}(U) \, \right) \right) )
    \end{split}
\end{equation}

Here, $W_1 \in \mathbb{R}^{\text{C/2} \times \text{C}}$ and $W_2 \in \mathbb{R}^{\text{C} \times \text{C/2}}$ are the weights of the fully connected layers of the \ac{mlp}. A different flavour of \textit{channel attention} was proposed by Woo \etal{} \cite{woo2018cbam}. Here, both global max pooling and global average pooling layers are included to generate two $C$-dimensional descriptors. The final \textit{channel attention} map $\alpha_{\mathbf{C}}$ is (Eq.~\ref{eq:attention_channel2}):

\begin{equation}\label{eq:attention_channel2}
\begin{split}
    \alpha_{\mathbf{C}} = \sigma ( \, & W_2 \, \text{ReLU} \left( W_1 \, \text{GlobalAvgPool}(U) \right) \, + \\
    & W_2 \, \text{ReLU} \left( W_1 \, \text{GlobalMaxPool}(U) \right) \, )
\end{split}
\end{equation}
\\
\textbf{\textit{Spatial attention}} aims to extract important information in the image domain or across the spatial dimensions of a feature map. In \myetal{Woo} \cite{woo2018cbam}, the \textit{spatial attention} block first performs average and max pooling operations across the channels of the input $U$, generating two feature maps which are then concatenated. A convolutional layer is then applied to produce a 1-channel spatial map which, after passing through the sigmoid function, becomes the attention map ($\alpha_{\mathbf{S}}$)(Eq.~\ref{eq: attention_spatial}):
\begin{equation}\label{eq: attention_spatial}
    \alpha_{\mathbf{S}} = \sigma \left( f \left( [\text{AvgPool}(U) ; \text{MaxPool}(U)] \right) \right)
\end{equation}

where $[\cdot ; \cdot]$ represents channel-wise concatenation and $f$ is the convolutional layer. \\

\myetal{Oktay} \cite{oktay2018attention} introduced a different version of \textit{spatial attention} with Attention U-Net, where attention gates apply convolutions to both features from the encoder and the corresponding decoder and then fuse them together to create the attention map. Moreover, instead of simply concatenating the encoder and decoder feature maps as for U-Net skip connections, the authors first scaled the encoder features with the generated \textit{spatial attention}.\\
\textit{\textbf{Non-local attention}}, proposed by \myetal{Wang} \cite{wang2018non}, aims to capture long-range dependencies by computing interactions between any two positions in an image or feature map. Conversely, \textit{channel} or \textit{spatial attention} focuses mainly on local information, \textit{i.e.}, the pooling operation leads to loss of spatial information, while convolutional layers process neighbourhood information. \\

In \textit{non-local attention}, three parallel $1 \times 1$ convolutional operations ($\theta$, $\phi$ and $g$) are applied on the input $U$, obtaining three compressed feature maps, while a final $1 \times 1$ convolutional operation $f$ restores the initial number of channels. Introduced by \myetal{Lin} \cite{lin2013network}, the $1 \times 1$ convolutions act as a channel-wise pooling operator.
The non-local attention map $\alpha_{\mathbf{NL}}$ is obtained through the following operations:
\begin{equation}
    \alpha_{\mathbf{NL}} = \tilde{X} = f \underbrace{\left( \text{softmax} \, \underbrace{ \left( \theta(U) \otimes \phi(U) ^T \right)}_{\beta_1 \in \mathbb{R}^{\text{HW} \times \text{HW}}} \otimes \, g(U) \right)}_{\beta_2 \in \mathbb{R}^{\text{HW} \times \text{C}/2}} 
\end{equation}
where $T$ is the matrix transpose operation, $\otimes$ is the matrix multiplication operator, and $\tilde{X}$, $\beta_1$ and $\beta_2$ are shown in Fig.~\ref{Attention.png}.
Moreover, the outputs of the convolutional layers $\theta$, $\phi$ and $g$ are reshaped to allow for matrix multiplication, \textit{i.e.}, they become 2D matrices of shape $\text{HW} \times \text{C/2}$.
These steps are shown in the third panel of Fig.~\ref{Attention.png}. \\

Finally, \textbf{\textit{self-attention}} is a mechanism in deep learning, closely related to the concept of \textit{non-local attention} \cite{wang2018non}, commonly used in \ac{nlp} tasks, particularly in transformer-based architectures \cite{vaswani2017attention}.
\myetal{Dosovitskiy} \cite{dosovitskiy2020image} adapted the \textit{self-attention} model to image-based applications. \\


\clearpage

In architectures based on transformers \cite{dosovitskiy2020image,hatamizadeh2022unetr} which employ \textit{self-attention} modules, the initial step involves splitting the input data into a sequence of patches.
Subsequently, these patches undergo processing via a linear projection layer and are merged with positional encodings to incorporate spatial biasing within the patch sequence (see Fig.~\ref{Attention.png}).
The embedded patches are then passed through a ``transformer encoder'' which consists of alternating layers of \textit{multi-head self-attention} and \ac{mlp} blocks, as well as residual connections and normalisation layers.
More specifically, a \textit{multi-head self-attention} block is composed of multiple parallel \textit{self-attention} heads, which compute attention scores based on the query ($Q$), key ($K$), and value ($V$) representations of the input as follows:
\begin{equation}
    \mathbf{SA} = \text{softmax} \left( \frac{Q K^T}{\sqrt{D_h}} \right) V
\end{equation}
where $D_h$ is a scaling factor. \\


\textbf{Advantages and disadvantages: }\textit{Attention}-based methods often add computational complexity to an existing \ac{dl} model, but have a differentiable objective and are easily trainable with gradient descent. 
Moreover, they aim to provide a weighting for inputs or internal features to focus the network on salient characteristics.
However, whether \textit{attention} can be regarded as feature importance is an ongoing debate \cite{jain2019attention,serrano2019attention,wiegreffe2019attention}.

\section{Applications}\label{sec:applications}

\subsection{Applications of perturbation-based methods (Table \ref{tab:perturbation_based_methods})} 

\subsubsection{Neurodegenerative disease classification}\label{sec:pert_nd}

\textit{Occlusion}, \textit{Swap Test} and \textit{Meaningful Perturbations} have all been applied to \ac{ad} classification networks trained on brain \ac{mri} images from the \ac{adni}\cite{tahmasebi2010validation}. Liu \etal{} \cite{liu2019deep} , Eitel \etal{} \cite{eitel2019testing}, and Yang \etal{} \cite{yang2018visual} employed \textit{Occlusion} to highlight image regions important for \ac{ad} prediction. Furthermore, Rieke \etal{} and and Yang \etal{} refined the \textit{Occlusion} method by occluding brain regions defined by an atlas, instead of image tiles \cite{rieke2018visualizing, yang2018visual}. Nigri \etal{} suggested \textit{Occlusion} may be unsuitable for neuroimaging data, since an occluded patch in a brain image from a cognitively normal individual may appear similar to disease \cite{nigri2020explainable}. Consequently, Nigri \etal{} proposed and applied \textit{Swap Test} \cite{nigri2020explainable}. \textit{Meaningful Perturbations} with a constant-valued mask was also applied for \ac{ad} classification \cite{thibeau2020visualization}. Furthermore, Shahamat \etal{} employed a complementary approach to \textit{Meaningful Perturbations}, in which minimal brain masks were learned for \ac{cnn}s trained to classify \ac{ad} and \ac{asd} \cite{shahamat2020brain}. A minimal brain mask \textit{keeps} the fewest brain regions whilst still achieving high model accuracy, whereas the brain mask in \textit{Meaningful Perturbations} \textit{deletes} the fewest brain regions that cause a wrong prediction. In most of these \ac{ad} studies, explanations contained salient regions known to be altered in AD such as the temporal lobe and hippocampus; however, the \textit{Occlusion} map in one study could not be meaningfully interpreted because the occlusion window was too large \cite{eitel2019testing}. \\

Another perturbation method, \textit{LIME}, was applied to explain predictions of a VGG \cite{simonyan2014very} model trained to predict \ac{pd} on \ac{spect} data  \cite{magesh2020explainable}. \textit{LIME} explanations of \ac{cn} individuals clearly delineated the putamen and caudate regions, whereas the highlighted areas in explanations for \ac{pd} patients often extended beyond these regions.\\

\subsubsection{Autism spectrum disorder classification}


Global perturbation-based methods have been employed to identify important features for \ac{asd} classification \cite{mellema2020architectural,li2018brain}. \textit{Permutation Feature Importance} was applied to a model trained on both structural (\textit{e.g.}, cortical volume and thickness) and functional \ac{mri} features \cite{mellema2020architectural}. Li \etal{} modified \textit{Occlusion} to produce a global explanation for \ac{asd} classification \cite{li2018brain}. Specifically, after a \ac{cnn} was trained to classify \ac{asd}, each atlas-based brain region $r$ was occluded in all images. Let $P_{\text{orig}}^{\text{\ac{cn}}}$ and $P_{\text{orig}}^{\text{\ac{asd}}}$ represent the distribution of predicted probabilities for all control and \ac{asd} subjects, respectively. Similarly, let $P_{\text{occ}}^{\text{\ac{cn}}}$ and $P_{\text{occ}}^{\text{\ac{asd}}}$ represent the distributions for the corresponding images with region $r$ occluded. Then, utilising Jensen-Shannon Divergence ($JSD$), the distance between class distributions was computed and compared (Eq.~\ref{eq:jsd}):

\begin{equation}\label{eq:jsd}
    \text{JSD}(P_{\text{orig}}^{\text{\ac{cn}}}, P_{\text{orig}}^{\text{\ac{asd}}}) > \text{JSD}(P_{\text{occ}}^{\text{\ac{cn}}}, P_{\text{occ}}^{\text{\ac{asd}}})
\end{equation}

Brain region $r$ was considered important if the decrease in JSD for occluding that region was statistically significant. The assumption is if region $r$ is important for \ac{asd} prediction, then the \ac{cnn} will not separate classes as effectively when region $r$ is removed. \\

Approaches that learn optimal brain masks for \ac{asd} classification have been used for models trained on the \ac{abide} \cite{di2014autism} I dataset \cite{dhurandhar2018explanations, shahamat2020brain}. Dhurandhar \etal{} produced the pertinent positive $\delta_{pos}$ and negative $\delta_{neg}$ features for a given resting-state \ac{fmri} image $I_0$ \cite{dhurandhar2018explanations}. The former, $\delta_{pos}$, are minimally sufficient meaning the network will predict the same class for both $I_0$ and $\delta_{pos}$. In contrast, $\delta_{neg}$ must be absent for the prediction i.e. the network predicts a different class for the perturbed image $I_0 + \delta_{neg}$ compared to $I_0$. Similarly, Shahamat \etal{} learned minimal brain masks for structural \ac{mri} images from \ac{abide} I, using the same approach they applied to the \ac{adni} dataset, previously described in \ref{sec:pert_nd} \cite{shahamat2020brain}. The important brain regions identified in all these \ac{asd} studies were disparate with little overlap, although motor regions were frequently highlighted. \\

\subsubsection{Schizophrenia classification}
A method analogous to \textit{Occlusion} was applied to a \ac{rnn} trained to classify \ac{scz} on \ac{rsfmri} data \cite{yan2019discriminating}. Initially, the data underwent dimensionality reduction using independent component analysis \cite{calhoun2001method}, and the time series of 50 valid \acp{ic} were retained. After \ac{rnn} training, feature importance of the $i^{th}$ \ac{ic} was computed by replacing the $i^{th}$ \ac{ic} time series values with its average (essentially occluding the $i^{th}$ feature), and then the change in model performance was assessed. The \acp{ic} with the greatest change in performance were considered to be the most important features for classification, which were located in the dorsal striatum and cerebellum.  \\

\subsubsection{Brain age regression}
The \textbf{\textit{U-noise}} method, similar to \textit{Meaningful Perturbations}, trained a U-net \cite{ronneberger2015u} by adding maximum random noise to input images without affecting the performance of a pre-trained prediction model \cite{unoise2021}. A sensitivity map can be generated to show the image pixels that were least tolerant to the addition of noise.  Bintsi \etal{} adapted the \textit{U-noise} architecture to interpret a 3D ResNet \cite{he2016deep} trained for brain age regression on UK Biobank \cite{sudlow2015uk} $T_1$-weighted \ac{mri} data \cite{bintsi2021voxel}. They computed an average importance map across all subjects for the test dataset. However, changes in the cerebral cortex related to aging were not well captured by the network. \\

\subsubsection{Sex classification}
\textit{Meaningful Perturbations} was compared to two alternative \ac{idl} methods for visualising model decisions for sex classification \cite{kan2020interpretation}. The \textit{Meaningful Perturbations} explanation highlighted regions of the frontal lobe, though the explanations were visually dissimilar between the three methods. \\

\begin{table}[h]
\footnotesize
\centering
\tabcolsep=0.11cm
\begin{tabular}{lcccl}
\hline
\textbf{Reference} &  \textbf{Data} & \textbf{Modality} & \textbf{\#Subjects} & \textbf{Method} \\
\hline \hline \\
\multicolumn{5}{l}{\textbf{AD Classification}} \\
\myetal{Eitel} \cite{eitel2019testing} & ADNI & sMRI (3D) & 344 & Occlusion \\
\myetal{Liu} \cite{liu2019deep} & ADNI & DTI (3D) & 151 & Occlusion \\ 
\myetal{Yang} \cite{yang2018visual} & ADNI & sMRI (3D) & 103 & Occlusion \\
 \myetal{Nigri} \cite{nigri2020explainable} & ADNI+AIBL & sMRI (3D) & 1,248 & Swap Test \\ 
\myetal{Thibeau} \cite{thibeau2020visualization} & ADNI+AIBL & sMRI (3D) & 1,171 & Meaningful Pert. \\ 
\myetal{Shahamat} \cite{shahamat2020brain} & ADNI & sMRI (3D) & 140 & Optimal mask \\ 
\myetal{Tang} \cite{tang2019interpretable} & In-house & Histology (2D) & 33 & Occlusion \\
\\

\multicolumn{5}{l}{\textbf{PD Classification}} \\
\myetal{Magesh} \cite{magesh2020explainable} &  PPMI & SPECT (2D) & 642 & LIME \\ 
\\

\multicolumn{5}{l}{\textbf{ASD Classification}} \\
\myetal{Li} \cite{li2018brain} & Various$^\dagger$ & fMRI (3D) & 225 & Occlusion (global) \\ 
\myetal{Dhurandhar} \cite{dhurandhar2018explanations} & ABIDE & fMRI (ts) & 293 & Optimal mask \\ 
\myetal{Shahamat} \cite{shahamat2020brain} & ABIDE & sMRI (3D) & 1,000 & Optimal mask \\ 
\myetal{Mellema} \cite{mellema2020architectural} & IMPAC & sMRI/fMRI (3D) & 915 & PFI \\   
\\

\multicolumn{5}{l}{\textbf{SCZ Classification}} \\
\myetal{Yan} \cite{yan2019discriminating} & In-house & fMRI (3D) & 1,100 & Leave-one-IC-out \\
\\

\multicolumn{5}{l}{\textbf{Sex Classification}} \\
 \myetal{Kan} \cite{kan2020interpretation} & HCP & sMRI (3D) & 1,113 & Meaningful Pert. \\ 
\\
\multicolumn{5}{l}{\textbf{Brain Age Regression}} \\
 \myetal{Bintsi} \cite{bintsi2021voxel} & UK Biobank & sMRI (3D) & 13,750 & U-Noise \\ 
\hline \hline
\end{tabular}
    \caption{\textbf{Articles using perturbation-based interpretable methods.} Abbreviations: ABIDE, Autism Brain Imaging Data Exchange; AD, Alzheimer's Disease; ADNI, Alzheimer’s Disease Neuroimaging Initiative; AIBL, Australian Imaging Biomarker and Lifestyle Flagship Study of Ageing; ASD, Autism Spectrum Disorder; DTI, Diffusion Tensor Imaging; fMRI, functional Magnetic Resonance Imaging; HCP, Human Connectome Project; IMPAC,  IMaging-PsychiAtry Challenge; LIME, local interpretable model-agnostic explanations; PD, Parkinson's Disease; PPMI, Parkinson’s Progressive Markers Initiative database; SCZ, schizophrenia spectrum disorders; sMRI, structural Magnetic Resonance Imaging; SPECT, Single Photon Emission Computed Tomography. Datasets: $\dagger$In-house+ABIDE}
    \label{tab:perturbation_based_methods}
\end{table}

\clearpage
\newpage
\subsection{Applications of gradient-based methods (Table \ref{tab:gradient_based_methods})} 

\subsubsection{Neurodegenerative disease classification}\label{sec:grad_nd}
Several studies employed \textit{Vanilla Gradients} or \textit{Grad $\times$ Input} to identify important brain regions after training a \ac{cnn} on \ac{mri} data from \ac{adni} for \ac{ad} classification \cite{oh2019classification, essemlali2020understanding, eitel2019testing, rieke2018visualizing}. All studies followed a similar approach, where class-average gradient maps were computed and then compared to explanations from other methods. Oh \etal{} produced \textit{Vanilla Gradients} maps for \ac{ad} classification and found they were in agreement with \textit{Occlusion} maps \cite{oh2019classification}. Essemlali \etal{} focused on connectivity between brain regions from diffusion-weighted \ac{mri} data, and found \textit{Vanilla Gradients} highlighted relevant brain regions, unlike \textit{Occlusion} \cite{essemlali2020understanding}. Eitel \etal{} analysed robustness across model runs for \textit{Grad $\times$ Input}, \textit{Occlusion} and two additional methods, and concluded \textit{Grad $\times$ Input} produced the least consistent explanations \cite{eitel2019testing}. To aid interpretation of \textit{Vanilla Gradients} maps, Rieke \etal{} computed a quantitative relevance score  by summing sensitivity within each atlas-based brain region \cite{rieke2018visualizing}. Eitel \etal{} similarly explored three summary statistics across brain regions, including mean region sensitivity to account for differences in brain region volume. All studies determined the medial temporal lobe and/or the hippocampus as important regions for \ac{ad} classification. \\

\subsubsection{Autism spectrum disorders classification}
\textit{Vanilla Gradients} has been applied to identify important features for \ac{asd} classification  from task-based \ac{fmri} data, where the task was testing perception of people's movements (biopoint task) \cite{li2019graph}. A graph was constructed with each node corresponding to a specific brain region, and having an associated feature vector of ten researcher-selected features. After training a \ac{gnn} and computing the gradient of the network output with respect to each feature, sensitivity maps were averaged across nodes and subjects to generate a sensitivity score per feature. \\

\subsubsection{Human immunodeficiency virus classification}
\textit{Vanilla Gradients} maps may include the influence of a confounding factor on the model decision, for example, patient age is often a confounding factor for neurodegeneration. Zhao \etal{} modified \textit{Vanilla Gradients} to remove the influence of age as a confounder from sensitivity maps computed from a \ac{cnn} classifier for \ac{hiv} \cite{zhao2019confounder}. Let $\text{f}_j=(f_{j1}, f_{j2}, ..., f_{jN})$ be the $j^{th}$ feature from the final convolutional layer for all $N$ subjects, $\textbf{s} =(s_1, ..., s_N)$ be \ac{cnn} score and $\textbf{a}=(a_1, ..., a_N)$ be subject age. Then the linear model $\text{f}_j = \beta_0 + \beta_1 \textbf{s} + \beta_2 \textbf{a}$ was fitted and if $\beta_2$ was non-zero, then age was defined as a confounder for the $j^{th}$ feature. When calculating the \textit{Vanilla Gradients} map, gradients were computed for unconfounded features only. The confounder-free sensitivity maps showed the posterior ventricle was most influenced by age rather than \ac{hiv}. \\

\subsubsection{Brain age regression}
\textit{SmoothGrad} was applied to a \ac{cnn} trained on a $T_1$-weighted brain \ac{mri} dataset to predict subject age, and a population-average sensitivity map was computed \cite{levakov2020deep}. The ventricles and subarachnoid cisterns were predominantly highlighted in the sensitivity map, which may be related to brain atrophy from the aging process. \\

\subsubsection{Cognitive task decoding}
McClure \etal{} trained a \ac{cnn} on activation maps from task-based \ac{fmri} to classify the \ac{fmri} task and employed \textit{Vanilla Gradients} for model interpretation \cite{mcclure2020evaluating}. To address the shattered gradients limitation of \textit{Vanilla Gradients} the authors utilised adversarial training. More specifically, non-targeted adversarial noise was learned and added to each image, optimised as the smallest magnitude noise that minimised the probability of the correct class. In this way, coupling \textit{Vanilla Gradients} with adversarial training was found to produce gradient maps that were more class discriminative than \textit{Vanilla Gradients}, \textit{Grad $\times$ Input}, and \textit{SmoothGrad}. However, the maps were still only found to be weakly correlated with class-specific features. \\

\Ac{lstm} is a deep learning architecture that is well-suited to time series \ac{fmri} data because it is designed to process sequence data. However, one limitation of applying \textit{Vanilla Gradients} to \ac{lstm} models is the issue of vanishing gradients \cite{bengio1994learning} when backpropagating through many timesteps. Consequently, only features in the latest time steps are highlighted in gradient maps. Ismail \etal{} proposed incorporating an attention mechanism into an \ac{lstm}, to bypass backpropagating through multiple timesteps during \textit{Vanilla Gradients} \cite{ismail2019input}. The attention-based \ac{lstm} was trained on task-based \ac{fmri} to classify the \ac{fmri} task, and \textit{Vanilla Gradients} was then able to highlight features in early time steps. \\

\begin{table}[h]
\footnotesize
\centering
\tabcolsep=0.11cm
\begin{tabular}{lcccl}
\hline
\textbf{Reference} &  \textbf{Data} & \textbf{Modality} & \textbf{\#Subjects} & \textbf{Method} \\
\hline \hline \\
\multicolumn{5}{l}{\textbf{AD classification}} \\
\myetal{Oh} \cite{oh2019classification} & ADNI & sMRI (3D) & 694 & Vanilla Gradients \\ 
\myetal{Eitel} \cite{eitel2019testing} & ADNI & sMRI (3D) & 344 & Grad $\times$ Input \\
\myetal{Essemlali} \cite{essemlali2020understanding} & ADNI & DW-MRI (2D) & 186 & Vanilla Gradients \\ \\

\multicolumn{5}{l}{\textbf{ASD classification}} \\
\myetal{Li} \cite{li2019graph} & In-house & fMRI (ts) & 118 & Vanilla Gradients \\ \\

\multicolumn{5}{l}{\textbf{HIV classification}} \\
\myetal{Zhao} \cite{zhao2019confounder} & In-house & sMRI (3D) & 355 & Vanilla Gradients \\ \\

\multicolumn{5}{l}{\textbf{Brain age regression}} \\
\myetal{Levakov} \cite{levakov2020deep} & Various$^{\dagger}$ & sMRI (3D) & 10,176 & SmoothGrad \\ \\

\multicolumn{5}{l}{\textbf{Cognitive task decoding}} \\
\myetal{McClure} \cite{mcclure2020evaluating} & HCP & fMRI (3D) & 965 & Vanilla Gradients \\
\myetal{Ismail} \cite{ismail2019input} & HCP & fMRI (ts) & 749 & Vanilla Gradients \\ 
\hline \hline
\end{tabular}
    \caption{\textbf{Articles using gradient-based methods.} Abbreviations: ABIDE, Autism Brain Imaging Data Exchange; AD, Alzheimer's Disease; ADNI, Alzheimer's Disease Neuroimaging Initiative; AIBL, Australian Imaging Biomarker and Lifestyle Flagship Study of Ageing; ASD, Autism Spectrum Disorder; DW-MRI, Diffusion-Weighted Magnetic Resonance Imaging; fMRI, functional Magnetic Resonance Imaging; HCP, Human Connectome Project; HIV, Human Immunodeficiency Virus; IXI, Information Extraction from Images; sMRI, structural Magnetic Resonance Imaging; ts, time series. Datasets: $\dagger$=ABIDE+ADNI+AIBL+IXI+others}
    \label{tab:gradient_based_methods}
\end{table}
\subsection{Applications of backpropagation-based methods (Table \ref{tab:backpropagation_based_methods})}

\subsubsection{Alzheimer's disease classification}
\textit{\ac{lrp}} and \textit{Guided Backpropagation} were utilised for \ac{ad} classification after training on \ac{adni} structural \ac{mri} data \cite{eitel2019testing,bohle2019layer}. In one study, \ac{lrp} heatmaps were shown to be more class-discriminative than \textit{Guided Backpropagation} maps \cite{bohle2019layer}. In a similar approach to the Eitel \etal{} study \cite{eitel2019testing}, the heatmap analysis was improved by three summary statistics: sum of relevance, mean relevance (to account for brain region size) and relevance gain compared to \ac{cn} (to find regions where explanations between \ac{ad} and \ac{cn} differ the most). All three studies consistently identified the hippocampi and other structures in the temporal lobe as important for \ac{ad} classification. \\
 
\subsubsection{Cognitive task decoding}
 \textit{\ac{lrp}-$\epsilon$} has been coupled with a deep learning model trained on task-based fMRI data to predict one of four cognitive states associated with viewing four image categories (body, face, place, or tool) during the task \cite{thomas2019analyzing}. The population-level explanation for each cognitive state was compared against a meta-analysis associated with the keyword from NeuroSynth. The explanations generally matched with the meta-analysis for body and face cognitive states, but less so for place and tool. \\

\begin{table}[!h]
\footnotesize
\centering
\tabcolsep=0.11cm
\begin{tabular}{lcccl}
\hline
\textbf{Reference} &  \textbf{Data} & \textbf{Modality} & \textbf{\#Subjects} & \textbf{Method} \\
\hline \hline \\

\multicolumn{5}{l}{\textbf{AD classification}} \\
\myetal{Bohle} \cite{bohle2019layer} & ADNI & sMRI (3D) & 344 & \ac{lrp}-$\alpha\beta$ \\
\myetal{Eitel} \cite{eitel2019testing} & ADNI & sMRI (3D) & 344 & Guided backprop + \ac{lrp} \\ \\
\multicolumn{5}{l}{\textbf{Sex classification}} \\
\myetal{Kan} \cite{kan2020interpretation} & HCP & sMRI (3D) & 1,113 & Guided backprop \\ \\
\multicolumn{5}{l}{\textbf{Cognitive task decoding}} \\
\myetal{Thomas} \cite{thomas2019analyzing} & HCP & fMRI (ts) & 100 & \ac{lrp}-$\epsilon$ \\ 
\hline \hline
\end{tabular}
\caption{\textbf{Articles using backpropagation-based methods}. Abbreviations: Alzheimer's Disease Neuroimaging Initiative; HCP, Human Connectome Project; LRP, Layer-wise Relevance Propagation; sMRI, structural Magnetic Resonance Imaging; fMRI, functional magnetic resonance imaging; ts, time series. Note: where \ac{lrp} variant not provided in paper, refer to as '\ac{lrp}'.}
\label{tab:backpropagation_based_methods}
\end{table}
\clearpage
\subsection{Applications of Class Activation Maps (Table \ref{tab:cam_methods})}

\subsubsection{Neurodegenerative disease classification}
\textit{CAM} and \textit{Grad-CAM} have been applied to \ac{ad} and \ac{mci} classification using a ResNet for 2D \ac{mri} \cite{khan2019transfer}, VGG and ResNet for 3D \ac{mri} \cite{zhang2021explainable, yang2018visual}, and \ac{gcn} for surface meshes of the cortex and sub-cortical structures \cite{azcona2020interpretation}, all trained on \ac{adni} structural \ac{mri} data. Additionally, Tang \etal{} applied \textit{Guided Grad-CAM} to amyloid-beta (A$\beta$) plaque-stained immunohistochemical data to classify plaque morphology, since A$\beta$ plaques are a histopathological hallmark of \ac{ad} \cite{tang2019interpretable}. \\ In their study of AD classification using a VGG, Zhang \etal{} showed that applying \textit{Grad-CAM} to lower convolutional layers produced more detailed explanations \cite{zhang2021explainable}. However, as lower layers tend to respond to edges/junctions of the brain images, so did the corresponding \textit{Grad-CAM} maps. \\

A related application of network interpretation is to use it to diagnose failure cases, for example, Khan \etal{} evaluated \textit{CAM} for a case of failed classification \cite{khan2019transfer}. In this case, the network attended to structures that are not associated with \ac{ad} classification, such as the skull. Similarly, Williamson \etal{} identified failure cases using \textit{Grad-CAM} for \ac{pd} classification of \ac{spect} scans, where the presence of noise artifacts and hyperintensities were shown to influence the network decision \cite{williamson2022improving}. \\

\subsubsection{Intracerebral hemorrhage classification}
\textit{Grad-CAM} generated explanations for a \ac{dl} model that detected and classified \ac{ich} sub-types from \ac{ct} scans of the head \cite{lee2019explainable}.  Ground truth data was available to validate the explanations; the proportion of ``bleeding points", selected by neuroradiologists to indicate the centre of haemorrhagic lesions, overlapping \textit{Grad-CAM} heatmaps was 78\%. \\

\subsubsection{Brain tumour classification}
Windisch \etal{} used \textit{Grad-CAM} to explain tumor classification from structural and diffusion MRI data \cite{windisch2020implementation}. As in the Khan \etal{} study \cite{khan2019transfer}, results were visually evaluated for cases of correct and incorrect classification. The network focused on the tumor when correctly classifying the scans, whilst there was no clear attention pattern when the classification failed. \\

\subsubsection{Autism spectrum disorder classification}
Li \etal{} proposed visualisation of activation maps from a \ac{gcn} trained to predict \ac{asd} from task-based \ac{fmri} data \cite{li2020pooling}. Using an approach analogous to \textit{CAM} but for \acp{gcn}, the 25\% of graph nodes (representing 21 brain regions) with the highest activation scores after the final graph-convolutional layer were visualised to interpret model classification. The method highlighted the dorsal striatum, thalamus, and frontal gyrus, regions thought to be affected by \ac{asd}. \\

\subsubsection{Sex classification}
The Dense-CAM network proposed by Gao \etal{} employed \textit{CAM} in the final layer of a DenseNet \cite{huang2017densely} trained for sex classification \cite{gao2019dense}, and found the cerebellum to be the most important brain region. On the other hand, Kim \etal{} \cite{kim2020understanding} applied \textit{Grad-CAM} to a \ac{gcn} trained on \ac{rsfmri} data and found regions involved in the default mode network to be important for sex classification, but not the cerebellum. \\ 

\subsubsection{Cognitive score prediction}
In two studies, \textit{Grad-CAM} was applied to visualise important brain regions for predicting the \ac{wrat} score of healthy individuals from the \ac{pnc}, using a regression \cite{qu2021ensemble} and classification approach \cite{hu2021interpretable}. In one study, a \ac{gcn} was trained on task-based \ac{fmri} data, then \textit{Grad-CAM} adapted for regression was computed \cite{qu2021ensemble}. Alternatively, subjects were classified into low, medium, and high \ac{wrat} score, and \textit{Guided Grad-CAM} was used to 
 identify important brain regions for \ac{wrat} score classification \cite{hu2021interpretable}. Both studies identified regions of the occipital lobe as important, which is involved in object recognition. \\

\begin{table}[!h]
\footnotesize
\centering
\tabcolsep=0.11cm
\begin{tabular}{lcccl}
\hline
\textbf{Reference} & \textbf{Data} & \textbf{Modality} & \textbf{\#Subjects} & \textbf{Method} \\
\hline \hline \\
\multicolumn{5}{l}{\textbf{AD classification}} \\
\myetal{Zhang} \cite{zhang2021explainable} & ADNI & sMRI (3D) & 1,407 & Grad-CAM \\ 
\myetal{Yang} \cite{yang2018visual} & ADNI & sMRI (3D) & 103 & CAM + Grad-CAM \\
\myetal{Azcona} \cite{azcona2020interpretation} & ADNI & Surface mesh (3D) & 435 & Grad-CAM \\ 
\myetal{Khan} \cite{khan2019transfer} & ADNI & sMRI (2D) & 150 & CAM \\ 
\myetal{Tang} \cite{tang2019interpretable}$^{\dagger}$ & In-house & Histology (2D) & 33 & Guided Grad-CAM \\ \\ 

\multicolumn{5}{l}{\textbf{PD classification}} \\
\myetal{Williamson} \cite{williamson2022improving} & PPMI & SPECT (3D) & 600 & Grad-CAM \\ 
\\
\multicolumn{5}{l}{\textbf{ASD classification}} \\
\myetal{Li} \cite{li2020pooling} & In-house & fMRI (ts) & 118 & Activation maps \\ 
\\
\multicolumn{5}{l}{\textbf{Tumour classification}} \\
 \myetal{Windisch} \cite{windisch2020implementation} & Various$^{\ddagger}$ & sMRI (2D) & 2, 479 & Grad-CAM \\ 
\\
\multicolumn{5}{l}{\textbf{Tumour segmentation}} \\
\myetal{Natekar} \cite{natekar2020demystifying} & BraTS & sMRI (3D) & 461 & Grad-CAM \\ 
\\
\multicolumn{5}{l}{\textbf{ICH classification}} \\
\myetal{Lee} \cite{lee2019explainable} & In-house & CT (2D) & 904 & CAM \\ 
\\
\multicolumn{5}{l}{\textbf{Sex classification}} \\
\myetal{Gao} \cite{gao2019dense} & Various$^{\dagger\dagger}$ & sMRI (3D) & 6,008 & CAM \\ 
\myetal{Kim} \cite{kim2020understanding} & HCP & fMRI (ts) & 1,094 & CAM + Grad-CAM \\ 
\myetal{Kan} \cite{kan2020interpretation} & HCP & sMRI (3D) & 1,113 & Grad-CAM \\ 
\\
\multicolumn{5}{l}{\textbf{Cognitive score prediction}} \\
\myetal{Hu} \cite{hu2021interpretable} & PNC & fMRI (ts) & 854 & Guided Grad-CAM \\ 
\myetal{Qu} \cite{qu2021ensemble} & PNC & fMRI (ts) & 800 & Grad-RAM \\ 
\hline \hline
\end{tabular}
\caption{\textbf{Articles using CAM interpretable methods.} Abbreviations: AD, Alzheimer's Disease; ADNI, Alzheimer's Disease Neuroimaging Initiative; ASD, Autism Spectrum Disorder; BraTS, Brain Tumor Segmentation challenge; CAM, Class Activation Map; CoRR, Consortium for Reliability and Reproducibility; CT, Computed Tomography; FCP, Functional Connectome Project; fMRI, functional Magnetic Resonance Imaging; GSP, Brain Genomics Superstruct Project; HCP, Human Connectome Project; ICH, Intracerebral Hemorrhage; IXI, Information Extraction from Images; NKI-RS, Nathan Kline Institute-Rockland Sample; PD, Parkinson's Disease; PNC, Philadelphia Neurodevelopmental Cohort; PPMI, Parkinson’s Progressive Markers Initiative database; RAM, Recurrent Activation Map; fMRI, functional Magnetic Resonance Imaging; SLIM, Southwest University Longitudinal Imaging Multimodal; sMRI, structural Magnetic Resonance Imaging; SPECT, Single Photon Emission Computed Tomography; TCGA, The Cancer Genome Atlas; ts, time series. $\dagger$= A$\beta$ plaque morphology classification. $\ddagger$= IXI, CyberKnife, TCGA. $\dagger\dagger$= HCP, FCP, GSP, NKI-RS, CoRR, SLIM.}
\label{tab:cam_methods}
\end{table}
\clearpage

\subsection{Applications of weight analysis (Table \ref{tab:weight_analysis_methods})}

\subsubsection{Tumor segmentation}

The  \textit{Concept Graphs} framework was applied to a U-Net brain tumor segmentation model trained on the \ac{brats} dataset \cite{bakas2017advancing}. The method identified multiple concepts at various model layers, such as the whole tumour, tumour core boundaries, and the tumour core region. Concept detection was also used for interpretability of a U-Net tumor segmentation model in Natekar \etal{} using \textit{Network Dissection}  \cite{natekar2020demystifying}. Results showed individual filters learned interpretable concepts including grey and white matter and edema; and separate filters for the whole tumor and the tumor core. These results showed that segmentation networks exhibit a modularity in the inference process that can be understood by humans. In Kori \etal{} \cite{kori2020abstracting}, in collaboration with a radiologist, inference trails that represent the trail of information in the network were also analysed. The network was shown to take a hierarchical approach to segmentation, starting with the detection of edges at lower layers and moving to the detection of the tumor in upper layers. \\

\subsubsection{ASD classification}

Dvornek \etal{} incorporated \textit{Community Detection} within their \ac{dl} model trained for \ac{asd} classification on \ac{rsfmri} data \cite{dvornek2019jointly}. The weights $W \in \mathbb{R}^{N \times K}$ for \textit{Community Detection}, where $w_{nk} \in W$ represents the strength of the connection between brain region $n$ and community $k$, were learned as part of the model. Clustering was then performed for each $k^{th}$ community vector $[w_{1k}, w_{2k}, ...,w_{Nk}]$ to assign each brain region as belonging \textit{vs.} not belonging to community $k$. Finally, the importance of community $k$ for \ac{asd} classification was defined as the sum of absolute weights of all $k$-indexed nodes in the classification model. The three most important communities included brain regions associated with language and social processing, memory, and reward-processing and decision-making. \\

\subsubsection{Cognitive task decoding}
\Acp{gcn} were re-designed for \textit{Community Detection} in BrainGNN \cite{li2021braingnn}. Let $W_i^{(l)}$ denote the learnable weights associated with node $i$ in graph convolutional layer $l$ of a GNN, where node $i$ represents a fixed brain region with one-hot location encoding $n_i$. The authors proposed to encode brain region location in $W_i^{(l)}$ by training a \ac{mlp} on the brain region location $n_i$:

\begin{equation}
vec(W_i^{(l)}) = \Theta_2^{(l)}\text{\ac{relu}}(\Theta_1^{(l)}n_i) + b^{(l)}
\end{equation}

where $\Theta_1^{(l)}$, $\Theta_2^{(l)}$ and $b^{(l)}$ are \ac{mlp} parameters. The elements $(\alpha_{nk})^+$ of $\text{\ac{relu}}(\Theta_1^{(l)}) \in \mathbb{R}^{N \times K}$ were interpreted as the non-negative community detection scores of brain region $n$ belonging to community $k$. In this study, BrainGNN was trained on the biopoint task-based \ac{fmri} dataset for \ac{asd} classification, as well as on the \ac{hcp} task-based \ac{fmri} data to classify seven cognitive tasks. \\

\begin{table}[!h]
\footnotesize
\centering
\tabcolsep=0.11cm
\begin{tabular}{lcccl}
\hline
\textbf{Reference} &  \textbf{Data} & \textbf{Modality} & \textbf{\#Subjects} & \textbf{Method} \\
\hline \hline \\

\multicolumn{5}{l}{\textbf{ASD classification}} \\
\myetal{Dvornek} \cite{dvornek2019jointly} & ABIDE & fMRI (ts) & 527 & Community detection \\ 
\myetal{Li} \cite{li2021braingnn} & Biopoint & fMRI (ts) & 115 & Community detection \\ 
\\
\multicolumn{5}{l}{\textbf{Tumour segmentation}} \\
\myetal{Kori} \cite{kori2020abstracting} & BraTS & sMRI (3D) & 300 & Concept Graphs \\ 
\myetal{Natekar} \cite{natekar2020demystifying} & BraTS & sMRI (2D) & 461 & Network Dissection \\ 
\\
\multicolumn{5}{l}{\textbf{Cognitive task decoding}} \\
\myetal{Li} \cite{li2021braingnn} & HCP & fMRI (ts) & 237 & Community detection \\ 

\hline \hline
\end{tabular}
\caption{\textbf{Articles using weight analysis interpretable methods.} Abbreviations: ABIDE, Autism Brain Imaging Data Exchange; ASD, Autism Spectrum Disorder; BraTS, Brain Tumor Segmentation challenge; fMRI, functional Magnetic Resonance Imaging; HCP, Human Connectome Project; sMRI, structural Magnetic Resonance Imaging.}
\label{tab:weight_analysis_methods}
\end{table}

\subsection{Applications of disentangled latent spaces (Table \ref{tab:disentangled_latent_space_methods})}

\subsubsection{Tumour classification}
A \textit{Capsule Network} was trained to classify tumour type (meningioma, pituitary, glioma) from segmented brain \acp{mri} \cite{afshar2018brain}. The \textit{Capsule Network} learned to reconstruct input images, where the latent space was constrained to three activation vectors representing the three tumour types. The activation vectors were inspected by perturbing individual vector elements and visualising the reconstructed images, which revealed the \textit{Capsule Network} had learned interpretable tumour features such as tumour size and elongation. \\

\subsubsection{Image generation}
One application of latent space disentanglement is training an \ac{ae} where a latent factor represents age, enabling the generation of \acp{mri} for different ages.\cite{zhao2019variational, mouches2021unifying}. One study jointly trained a supervised age regression network and a \ac{vae} where both shared convolutional layers \cite{zhao2019variational}. The latent space of the \ac{vae} was trained to approximate a prior distribution $p(z | \hat{y})$ conditioned on the age $\hat{y}$ predicted by the regressor. In another study, an \ac{ae} was coupled with a linear function such that the first parameter of the latent space predicted subject age \cite{mouches2021unifying}. Age-specific \ac{mri} images were synthesised in both studies by adjusting the age-related latent factor. \\
One study synthesised T1w \ac{mri} images at different ages for healthy controls and for patients with \ac{ad}, through disentangling the effect of \ac{ad} from healthy ageing on \acp{mri}during image reconstruction \cite{ouyang2022disentangling}. Disentanglement was achieved by learning two orthogonal directions in the latent space of an \ac{ae} and designing loss functions that encouraged the difference between two successive \acp{mri} of a subject, as represented in the latent space, to be equal to the sum of two components in the healthy and diseased directions. Formally, let $z_1$ and $z_2$ be the latent representations of two \ac{mri} images acquired for a subject at times $t_1$ and $t_2$, respectively, with $t_2 > t_1$. Then the vector $\Delta z = (z_2 - z_1) / (t_2 - t_1)$ was constrained by the loss so that $\Delta z \approx \Delta z_a + \Delta z_d$, where $\Delta z_a$ and $\Delta z_d$ are parallel to the healthy and disease directions, respectively. The disentangled latent factors were visualised and the \ac{ae} had learned distinct trajectories for \ac{cn}, \ac{pmci}, \ac{smci} and \ac{ad} subjects. \\

Another application is harmonisation of \ac{mri} across different clinical sites, where the contrast of an \ac{mri} image acquired at site A is transformed to appear as if it were acquired at site B, whilst leaving the subject anatomy unaltered. Disentangled latent spaces have been employed in \ac{mri} harmonisation models to learn separate features for anatomy \textit{vs.} image contrast \cite{zuo2021unsupervised, zhao2023disentangling}. For example, Zuo \etal{} trained a model named CALAMITI for site harmonisation across 10 clinical sites \cite{zuo2021unsupervised}. Two separate encoders, $E_{anat}$ and $E_{cont}$, learned feature representations for anatomical and contrast information during image reconstruction, respectively $\theta_{anat}$ and $\theta_{cont}$. $E_{anat}$ learned anatomical-only features by encouraging $\theta_{anat}$ from both T1-w and T2-w images of the same subject and slice number to be identical, thus capturing the shared anatomical features and ignoring contrast. Simultaneously, $E_{cont}$ was trained on different slice numbers of the same T1-w and T2-w images to represent the residual features of contrast in $\theta_{cont}$, after accounting for anatomy in $\theta_{anat}$. Similarly, Zhao \etal{} learned a disentangled \ac{vae} latent space consisting of site-related \textit{vs.} non site-related features \cite{zhao2023disentangling}. For $M$ sites, a vector of length $M$ in the latent space was optimised to represent site-specific features by feeding it into a site classification network and minimising the cross-entropy loss. Both studies visualised the latent space and demonstrated subjects were clustered by site. \\

\subsubsection{Brain age regression}

Disentanglement of a network latent space may be advantageous for networks trained on multi-modal data, to uncouple modality-specific from shared latent features \cite{hu2020disentangled}. Such an approach was adopted when predicting infant brain age from both \ac{fmri} and \ac{smri} features, where an \ac{ae} was trained on each modality such that the latent space was divided into modality-specific features and common features \cite{hu2020disentangled}. The common feature vectors from both \acp{ae} for the same subject were encouraged to be identical by adding an $L_2$ loss and adversarial loss. Furthermore, each decoder was required to reconstruct the input data from its own common feature vector as well as that from the \ac{ae} of the other modality, a method known as cross reconstruction. The common feature vector and each modality-specific feature vector were passed to an age prediction network to predict age. Visualisation of the learned latent space demonstrated the latent features were ordered by age. \\

\subsubsection{Neurodegenerative disease classification}

Similarly, Wang \etal{} learned a disentangled latent space to partition features by modality, as well as learn features shared between modalities, when learning from multimodal data \cite{wang2023deep}. Two \acp{ae} were trained, one on each modality, on \ac{smri}/\ac{fdgpet} for \ac{ad} classification from \ac{adni}, as well as on \ac{smri}/\ac{dti} for \ac{pd} classification from \ac{ppmi}. Each latent space was split into modality-specific and common features; a distance loss, defined as  the $L_2$ loss between common features divided by the $L_2$ loss between modality-specific features, encouraged common features to be identical and modality-specific features to be dissimilar. In addition, cross-reconstruction was adopted, where each decoder generated images using common features learned from both \acp{ae}. Disentanglement enabled important brain regions to be identified as \ac{smri}-specific, \ac{fdgpet} (or \ac{dti} in the case of \ac{pd} classification)-specific or common to both imaging modalities. \\

\begin{table}[h]
\footnotesize
\centering
\tabcolsep=0.11cm
\begin{tabular}{lcccl}
\hline
\textbf{Reference} &  \textbf{Data} & \textbf{Modality} & \textbf{\#Subjects} & \textbf{Method} \\
\hline \hline \\
\multicolumn{5}{l}{\textbf{Image generation}} \\
\myetal{Zhao} \cite{zhao2019variational} & In-house & sMRI (3D) & 245 & Factor: age \\ 
\myetal{Mouches} \cite{mouches2021unifying} & In-house & sMRI (3D) & 2,681 & Factor: age \\
& +IXI & & & \\
\myetal{Zuo} \cite{zuo2021unsupervised} & Various$^\dagger$ & sMRI (2D) & 100 & Factor: clinical site \\
\myetal{Zhao} \cite{zhao2023disentangling} & Various$^\ddagger$ & Surface mesh (3D) & 2,542 & Factor: clinical site \\ 
\myetal{Ouyang} \cite{ouyang2022disentangling} & ADNI & sMRI (3D) & 632 & Factor: healthy+disease \\
\\
\multicolumn{5}{l}{\textbf{Tumour classification}} \\
\myetal{Afshar} \cite{afshar2018brain} & In-house & sMRI (2D) & 233 & Capsule network \\ \\ 

\multicolumn{5}{l}{\textbf{Brain age regression}} \\
\myetal{Hu} \cite{hu2020disentangled} & BCP & sMRI+fMRI (3D) & 178 & Factor: modality \\ \\ 

\multicolumn{5}{l}{\textbf{Neurodegenerative disease classification}} \\
\myetal{Wang} \cite{wang2023deep} & \phantom{+}ADNI & sMRI+PET (3D) & 4,126 & Factor: modality \\ 
& +PPMI & DTI (3D) & & \\
\hline \hline

\end{tabular}
\caption{\textbf {Articles using disentangled latent space methods.} Abbreviations: ADNI, Alzheimer's Disease Neuroimaging Initiative; BCP, UNC/UMN Baby Connectome Project; BLSA, Baltimore Longitudinal Study of Aging; BraTS, Brain Tumor Segmentation Challenge; fMRI, Functional Magnetic Resonance Imaging; IBIS, Infant Brain Imaging Study; IXI, Information Extraction from Images; OASIS, Open Access Series of Imaging Studies; sMRI, Structural Magnetic Resonance Imaging. Datasets: $\dagger$=IXI+OASIS+BLSA, $\ddagger$=In-house+BCP+IBIS}
    \label{tab:disentangled_latent_space_methods}
\end{table}

\subsection{Applications of interpretable hybrid models (Table \ref{tab:hybrid_model_methods})}

\subsubsection{Neurodegenerative disease classification}

One blueprint for interpretable intermediate features for \ac{ad} classification is a heatmap of predicted probabilities of \ac{ad} across brain regions. Qiu \etal{} adopted this approach when designing a \ac{nn} + \ac{nn} hybrid model, where the first component was a patch-based \ac{cnn} that output probability of \ac{ad} across brain patches \cite{qiu2020development}. After training, predicted probabilities for 200 voxels were concatenated with non-imaging features (age, gender, \ac{mmse}) and used to train a multi-layer perceptron to predict \ac{ad} (summarised in Fig. \ref{fig:hybrid}). However, the heatmaps were less precise and therefore more difficult to interpret than the next two studies discussed. Similarly,  a \ac{nn}+\ac{ml} hybrid model learned intermediate probability heatmaps for \ac{ad} \cite{lee2019toward}. For the first component, an ensemble of \ac{nn} classifiers was trained to predict \ac{ad} or \ac{mci} status for each of 93 atlas-derived brain regions, from which a probability heatmap was constructed. The second component was a linear \ac{svm} trained to predict \ac{ad} status from the probability heatmap. This study followed the ``int. features + feature importance" approach and also considered the weights learned by the linear \ac{svm}. Nguyen \etal{} adopted a comparable approach where instead of learning a brain heatmap of probabilities, voxels were assigned a grade close to 1 if abnormal and close to -1 if healthy \cite{nguyen2022interpretable}. The NN + NN hybrid model consisted of a set of patch-based U-Nets that generated the grade heatmap, followed by a \ac{gcn} trained to predict \ac{cn} \textit{vs.} \ac{ad} \textit{vs.} \ac{ftd}. The population-average heatmaps in all three studies were highly class-discriminative and were consistent with known disease pathology, focusing predominantly on the temporal lobe for \ac{ad}, and the frontal lobe for \ac{ftd} in the case of the study by Nguyen \etal{}. \\

Three studies employed \textit{prototypes} to introduce interpretability into \ac{ad} classification models \cite{mohammadjafari2021using, wolf2023don, mulyadi2023estimating}. ProtoPNet has been trained on two public T1-w \ac{mri} datasets,\ac{adni} and \ac{oasis} \cite{marcus2007open}, to classify \ac{ad} \cite{mohammadjafari2021using}. Wolf \etal{} trained a variant of ProtoPNet to predict \ac{ad} from \ac{fdgpet} images from the \ac{adni} database \cite{wolf2023don}, and two of the prototypes highlighted the ventricles and occipital lobe. In another study, a prototype-based model was trained on T1-w \ac{mri} from \ac{adni} and an in-house dataset, and the \textit{prototypes} learned were reconstructed from the latent space to a 3D feature vector [\ac{ad} diagnosis, \ac{mmse}, age] \cite{mulyadi2023estimating}. Furthermore, prototypical brains for each diagnosis class (\ac{cn}, progressive \ac{mci}, stable \ac{mci} and \ac{ad}) were compared to individual scans; for example, a \ac{cn} subject differed most from the \ac{ad} prototypical brain in the amygdalae. However, a limitation of prototype-based explanations is they are low-resolution because of upsampling from a latent space to the image space. \\

Abuhmed \etal{} explored yet another hybrid model design and predicted \ac{ad} clinical scores as interpretable intermediate features\cite{abuhmed2021robust}. The \ac{nn}+\ac{ml} hybrid model in the predicted prognosis of \ac{ad} status at month 48 from multi-modal data collected at baseline and months 6, 12, and 18. The first component was a Bi-\ac{lstm} model trained to forecast seven cognitive scores (such as the \ac{mmse}) at month 48. The second component was an interpretable \ac{ml} classifier trained separately to forecast disease status at month 48 from the seven forecasted cognitive scores, subject age, gender, and education. Taking the ``int. features + feature importance" approach, explanations were also generated from the \ac{ml} classifier; however, the explanations were only in relation to the cognitive scores and not the neuroimaging data. \\


\subsubsection{Autism spectrum disorder classification}
A \textit{prototype-based} approach was adopted when classifying \ac{asd} from \ac{fc} matrices computed from \ac{rsfmri} data from the \ac{abide} dataset \cite{kang2022prototype}. The first component of the hybrid model was a transformer network (see Section \ref{sec:intrinsic_attention}) that generated latent features for a subject, and then the predicted class was determined by the class prototype closest to the latent features. To enable interpretability, a decoder was trained to reconstruct the input \ac{fc} from an individual's latent features and was also used to decode a more prototypical \ac{fc}. For example, by subtracting a reconstructed \ac{fc} of a control subject from the \ac{asd}-typical version, the authors found regions such as the right cingulate gyrus and the occipital and frontal poles as the most different from \ac{asd} for this individual. \\

\subsubsection{Brain age regression}
\textit{Prototypes} were also employed for predicting brain age from T1-w \ac{mri} images[cite IXI], as well as predicted gestational age from fetal \ac{us} images \cite{hesse2023prototype}. \textit{Prototypes} were adapted for regression as follows: prototypes were not assigned to a class, but each prototype was replaced with the closest latent representation of a training image, and associated with the corresponding age label. The predicted age is the weighted mean of age labels of all prototypes within a fixed distance from the sample in the latent space. The method was able to display the four prototypical brains most similar to a test image. \\

\subsubsection{Attention deficit hyperactivity disorder classification}
Another choice for intermediate interpretable features in a hybrid model is to learn \acp{fbn} that are important for the \ac{dl} task. Qiang \etal{} trained such a \ac{ml} + \ac{ml} hybrid model to classify \ac{adhd} from \ac{rsfmri} data \cite{qiang2020deep}. First, a \ac{vae} was trained on \ac{fmri} data and the latent representations learned by the \ac{vae} were used to learn \ac{fbn} weights using Lasso regression (penalised regression with L1 penalty). For the second component, \ac{fc} matrices were constructed from the \ac{fbn} weights and used to train an \ac{ml} classifier to predict \ac{adhd}. The \acp{fbn} learned by the \ac{vae} were shown to be similar to those derived from another state-of-the-art method.  \\

\subsubsection{Cognitive/clinical score regression}
In a similar manner to Qiang \etal{}, D'Souza \etal{} learned \acp{fbn} as interpretable intermediate features for cognitive and clinical score prediction \cite{shimona2020deep}. The authors coupled representation learning with a \ac{nn} that predicted cognitive or clinical scores. Intermediate \acp{fbn} were learned from \ac{rsfmri} functional connectivity matrices using \ac{srddl}. Simultaneously, an \ac{lstm} was trained from the subject-specific \ac{fbn} coefficients. The \ac{ml} + \ac{nn} hybrid model was trained to predict the Cognitive Fluid Intelligence Score for healthy subjects from the \ac{hcp} dataset \cite{van2013w}, as well as clinical scores (Autism Diagnostic Observation Schedule, Social Responsiveness Scale and Praxis) for \ac{asd} subjects from the \ac{kki} dataset \cite{bigler2008neuropsychology}. Fifteen learned \acp{fbn} for both \ac{hcp} and \ac{kki} data were presented as the model explanations. For example, several of the networks learned to predict the Cognitive Fluid Intelligence Score were involved in the Medial Frontal Network and the Frontal Parietal Network, which play a role in decision-making, attention, and working memory. \\

\begin{table}[!h]
\footnotesize
\centering
\tabcolsep=0.11cm
\begin{tabular}{lcccl}
\hline
\textbf{Reference} &  \textbf{Data} & \textbf{Modality} & \textbf{\#Subjects} & \textbf{Method} \\
\hline \hline \\
\multicolumn{5}{l}{\textbf{AD classification}} \\
\myetal{Abuhmed} \cite{abuhmed2021robust} & ADNI & sMRI (3D) & 1,371 & Cognitive scores \\
 & & PET (3D) & & \\
\myetal{Lee} \cite{lee2019toward} & ADNI & sMRI (3D) & 801 & Disease \\
& & & &probability map \\
\myetal{Liu} \cite{liu2020masked} & ADNI & sMRI (3D) & 3,021 & Int. features only \\
\myetal{Nguyen} \cite{nguyen2022interpretable} & Various$^{\dagger}$ & sMRI (3D) & 2,036 & Disease grade map \\
\myetal{Qiu} \cite{qiu2020development} & Various$^{\dagger\dagger}$ & sMRI (3D) & 1,446 & Disease \\
& & & & probability map \\
\myetal{Mohammadjafari} \cite{mohammadjafari2021using} & \phantom{+}ADNI & sMRI (2D) & 408 & Prototypes \\
& +OASIS & & & \\
\myetal{Mulyadi} \cite{mulyadi2023estimating} & \phantom{+}ADNI & sMRI (3D) & 2,285 & Prototypes \\
& +In-house & & & \\
\myetal{Wolf} \cite{wolf2023don} & ADNI & PET (3D) & 1,245 & Prototypes \\ \\

\multicolumn{5}{l}{\textbf{ASD classification}} \\
\myetal{Kang} \cite{kang2022prototype} & ABIDE & fMRI (3D) & 985 & Prototypes \\ \\

\multicolumn{5}{l}{\textbf{Brain age regression}} \\
\myetal{Hesse} \cite{hesse2023prototype} & IXI & sMRI (2D) & 562 & Prototypes \\ \\

\multicolumn{5}{l}{\textbf{Cognitive/ clinical score regression}} \\
\myetal{D'Souza} \cite{shimona2020deep} & \phantom{+}HCP & fMRI (3D) & 150 & FBNs \\ 
 & +KKI & DTI (3D) & & \\ \\

\multicolumn{5}{l}{\textbf{ADHD classification}} \\
\myetal{Qiang} \cite{qiang2020deep} & ADHD-200 & fMRI (3D) & 541 & FBNs \\ 
\hline \hline 
\end{tabular}
\caption{\textbf{Articles using interpretable hybrid models or interpretable intermediate features}. Abbreviations: ABIDE, Autism Brain Imaging Data Exchange; AD, Alzheimer Disease; ADHD, Attention Deficit Hyperactivity Disorder; ADNI, Alzheimer’s Disease Neuroimaging Initiative; AIBL, Australian Imaging Biomarker and Lifestyle Flagship Study of Ageing; ASD, Autism Spectrum Disorder; DTI, Diffusion Tensor Imaging; FBN, Functional Brain Network, FHS, Framingham Heart Study; fMRI, functional Magnetic Resonance Imaging; HCP, Human Connectome Project; IXI, Information eXtraction from Images; KKI, Kennedy Krieger Institute; MIRIAD, Minimal Interval Resonance Imaging in Alzheimer’s Disease; NACC, National Alzheimer’s Coordinating Center; NIFD, Frontotemporal lobar Degeneration Neuroimaging Initiative; OASIS, Open Access Series of Imaging Studies; PET, Positron Emission Tomography; sMRI, structural Magnetic Resonance Imaging. $\dagger$=ADNI+AIBL+OASIS+MIRIAD+NIFD, $\dagger\dagger$=ADNI+AIBL+FHS+NACC.}
\label{tab:hybrid_model_methods}
\end{table}
\subsection{Applications of interpretable generative models (Table \ref{tab:generative_methods})}

\subsubsection{Neurodegenerative disease maps}

Several studies trained a \ac{gan} \cite{goodfellow2020generative} on the \ac{adni} structural \ac{mri} dataset to predict disease effect maps for \ac{ad}, considering either \ac{mci} or \ac{cn} as the control class \cite{baumgartner2018visual, lanfredi2020interpretation, liu2021going, bass2020icam, bass2022icam}. Baumgartner \etal{} developed a visual attribution method based on a conditional \ac{gan} (\textbf{\textit{VA-\ac{gan}}}) \cite{baumgartner2018visual}. In this work, an additive map $M(x)$ was learned as a function of an input image $x$ from the \ac{ad} class, such that the modified image $x + M(x)$ appears cognitively normal. In contrast to learning an additive map, Lanfredi \etal{} trained a \ac{gan} to generate a deformation field, known as deformation field interpretation (\textbf{\textit{DeFI-\ac{gan}}}), which was shown to produce sparser disease effect maps than VA-\ac{gan} \cite{lanfredi2020interpretation}. The deformation field transforms an image from the \ac{ad} class to the \ac{mci} class by modelling brain atrophy. As such, deformation-based approaches are only appropriate for modelling diseases where brain atrophy is the predominant imaging marker. The same deformation field approach was employed by Liu \etal{}, but using a cycle\ac{gan} that generated modified \ac{ad} and \ac{cn} images \cite{liu2021going}. The Jacobian of the deformation field was visualised as the disease effect map. \\

The aforementioned methods assume that the category labels of the test data (either real or estimated by a separate classifier) are known during testing, meaning that the models can generate explanations, but cannot perform the classification. Bass \etal{}\cite{bass2020icam} developed a model that both classified disease and generated disease effect maps. By incorporating a classification network, this model obviates the need for previously classified data during testing. A \ac{vae}-\ac{gan} was trained to disentangle class-relevant features from background features, and therefore to separate the effects of healthy aging from disease. The mean and variance of predicted disease effect maps were sampled from the latent space during testing, as opposed to from a single additive map for each subject. The method was applied to brain structural \ac{mri} data from \ac{adni} as in \cite{baumgartner2018visual} and disease effect maps were shown to improve when compared to \textit{VA-\ac{gan}} and gradient-based methods. In addition to classification, the method was extended for regression of \ac{mmse} from structural \ac{mri} \ac{adni} data; regression of age from Biobank brain structural MRI scans; and regression of birth age from \ac{dhcp} data \cite{bass2022icam}. All of these studies produced \ac{ad} disease effect maps that successfully modelled atrophy of the ventricles, hippocampus, and cortical grey matter known to occur in \ac{ad}. \\

\subsubsection{Brain tumour and stroke segmentation}

More recently, state-of-the-art diffusion models \cite{ho2020denoising} have been trained to predict disease effect maps (anomaly maps) for neuroimaging datasets \cite{bercea2023reversing, sanchez2022healthy, wolleb2022diffusion}. Two studies trained a \ac{ddpm} on \ac{smri} images from the \ac{brats} dataset to convert a cancerous MRI to appear healthy, and a third study trained a generative model to transform MRI brain images of stroke patients to appear healthy. In all studies, the generated healthy image was subtracted from the original to produce the anomaly map. Wolleb \etal{} trained an unconditional \ac{ddpm} and a classifier, and then used classifier guidance to transform an MRI from cancerous to healthy \cite{wolleb2022diffusion}. In contrast, Sanchez \etal{} trained a conditional \ac{ddpm} and employed classifier-free guidance to alter the cancerous images \cite{sanchez2022healthy}. Bercea \etal{} implemented a two-stage approach, where stroke-effected regions were removed from the image in stage one, and then stage two comprised an in-painting generative model to fill in these erased regions as healthy \cite{bercea2023reversing}. The anomaly maps in all three studies were shown to identify pathological brain regions successfully. \\

\begin{table}[h]
\footnotesize
\centering
\tabcolsep=0.11cm
\begin{tabular}{lcccl}
\hline
\textbf{Reference} &  \textbf{Data} & \textbf{Modality} & \textbf{\#Subjects} & \textbf{Method} \\
\hline \hline \\

\multicolumn{5}{l}{\textbf{AD classification}} \\
\myetal{Baumgartner} \cite{baumgartner2018visual} & ADNI & sMRI (3D) & 1,288 & Generative additive maps  \\ 
\myetal{Bass} \cite{bass2020icam}  & ADNI  & sMRI (3D) & 1,053 &  Generative additive maps \\ 
\myetal{Bass} \cite{bass2022icam}  & ADNI  & sMRI (3D)& 1,053  &  Generative additive maps \\ 
\myetal{Lanfredi} \cite{lanfredi2020interpretation} & ADNI & sMRI (3D) & 825 & Generative \\
& & & &deformation fields \\ 
\myetal{Liu} \cite{liu2021going} & ADNI & sMRI (3D) & 1,344 & Generative \\
& & & &deformation fields \\ 
\\

\multicolumn{5}{l}{\textbf{Brain age regression}} \\
\myetal{Bass} \cite{bass2022icam} & Various$^\dagger$ & sMRI (3D) & 12,434 & Generative additive maps \\ \\ 

\multicolumn{5}{l}{\textbf{Brain tumour and stroke segmentation}} \\
Bercea \myetal{} \cite{bercea2023reversing} & Various$^\ddagger$ & sMRI (2D) & 1,412 & Generative additive maps \\ 
Sanchez \myetal{} \cite{sanchez2022healthy} & BraTS & sMRI (2D)  & 1,251 & Generative additive maps \\ 
Wolleb \myetal{} \cite{wolleb2022diffusion} & BraTS & sMRI (2D) & N/A & Generative additive maps \\ 
\hline \hline
\end{tabular}
\caption{\textbf{Articles using generative models.} Abbreviations: 
 AD, Alzheimer's Disease; ADNI, Alzheimer's Disease Neuroimaging Initiative; BraTS, Brain Tumour Segmentation Challenge; dHCP, Developing Human Connectome Project;  IXI, Information eXtraction from Image; N/A, Not Available; sMRI, structural Magnetic Resonance Imaging. $\dagger$=UK Biobank+dHCP, $\ddagger$=ATLAS v2.0+IXI+FastMRI}
    \label{tab:generative_methods}
\end{table}

\subsection{Applications of deep structural causal models (Table \ref{tab:causal_model_methods} )}

Reinhold \etal{} extended the \textit{\ac{dscm}} in Eqn.~\ref{eq:pawlowski_dscm} to model causal effects for structural \ac{mri} images from a \ac{ms} cohort by adding duration of \ac{ms} symptoms, expanded disability severity score, lesion volume and image slice number \cite{reinhold2021structural}. Counterfactual difference maps were explored, such as the counterfactual $do(l = 0 \text{ mL})$ for a brain \ac{mri} of an \ac{ms} patient, where the model successfully removed the \ac{ms} lesions from the counterfactual image. \\




Furthermore, Rasal \etal{} modified a \ac{dscm} to synthesise 3D surface meshes of the brain stem by introducing graph convolutional layers into the VAE \cite{rasal2022deep}. The authors performed interventions on the population-mean brain stem, as well as generating subject-specific counterfactual surface meshes for variables such as age and sex. Realistic counterfactual meshes were generated for scenarios outside the true data distribution, for example, $do (\text{age} = 80 \text{year-old})$ when the maximum participant age was 70 years old. \\ 

\begin{table}[!h]
\footnotesize
\centering
\begin{tabular}{lcccl}
\hline
\textbf{Reference} &  \textbf{Data} & \textbf{Modality} & \textbf{\#Subjects} & \textbf{Method} \\
\hline \hline \\
\multicolumn{5}{l}{\textbf{Image generation}} \\
Pawlowski \myetal{} \cite{pawlowski2020deep} & UK Biobank & sMRI (2D) & 13,750 & DSCM \\ 
Reinhold \myetal{} \cite{reinhold2021structural} & In-house & sMRI (2D) & 77 & DSCM \\ 
Rasal \myetal{} \cite{rasal2022deep} & UK Biobank & Surface meshes (3D) & 14,502 & DSCM \\ 
\hline \hline
\end{tabular}
    \caption{\textbf{Articles using deep structural causal models.} Abbreviations: DSCM, Deep Structural Causal Model; sMRI, structural Magnetic Resonance Imaging.}
    \label{tab:causal_model_methods}
\end{table}

\subsection{Applications of attention mechanisms (Table \ref{tab:attention_based_methods})}

\subsubsection{Image segmentation}
\myetal{Gu} \cite{gu2020net} introduced \textit{channel}, \textit{spatial}, and \textit{non-local attention} blocks in a modified U-Net to improve the performance of medical image segmentation tasks. More specifically, they used \textit{spatial attention} blocks throughout the decoder layers of the U-Net by combining both higher (from the decoder) and lower resolution (from the encoder) feature maps, similar to that proposed previously \cite{oktay2018attention}.
\textit{Channel attention} blocks were also introduced after each decoding layer by global average pooling and global max pooling \cite{woo2018cbam}. The latter was also introduced as ``scale attention'', which assigns a weight for each of the decoder outputs to enable differential attention to be assigned to a given input. The final \textit{non-local block} was introduced at the lowest resolution level (the bottleneck of the U-Net) due to its complexity.
They showed the \textit{spatial attention} maps from the trained network were able to highlight the object to be segmented, suggesting that the use of \textit{attention} enhanced the ability of the network to focus on target areas to facilitate performance.

\subsubsection{Disease classification}
A 3D \textit{spatial attention} network was used to classify \ac{ad} using two large structural \ac{mri} datasets (\ac{adni} and an in-house database) \cite{jin2020generalizable}.
Following grey matter segmentation, volumes were inputted into a 3D-\ac{cnn}, which contained a \textit{spatial attention} block after the first three convolutional layers to highlight important regions in the feature maps.
However, the \textit{spatial attention} module contained a ReLU rather than a sigmoid activation function.
Thus, probability values for each spatial location were not produced, but nevertheless, the method was able to identify those brain regions correlated with atrophy, characteristic of \ac{ad}. \\

\textit{Attention} has also been introduced into a hybrid \ac{dl} framework to classify \ac{scz} and \ac{asd} using an in-house and the ABIDE \ac{rsfmri} dataset, respectively \cite{zhao2022attention}. Features were first extracted from the imaging data using principal components analysis, and 50 independent components (IC) were retained per subject, each of which was a times series. An attention-guided \ac{crnn} was then used to process the \ac{ic} time series data, and a \ac{dnn} for processing \ac{fnc} matrices.
The \ac{crnn} \textit{attention} block aimed to highlight which \ac{rsfmri}-derived \acp{ic} were more significant for prediction.
The \textit{attention} module was comparable to that proposed by \myetal{Woo} \cite{woo2018cbam}, which uses both max and average pooling layers, but \myetal{Zhao} \cite{zhao2022attention} applied these along the time axis.
The outputs of these two separate networks were concatenated and passed through a logistic regressor to obtain the final classification result. \\
\myetal{Sarraf} \cite{sarraf2023ovitad} developed an optimized vision transformer, OViTAD, for classifying \ac{hc}, \ac{mci}, and \ac{ad} brains using \ac{rsfmri} and structural \ac{mri} data. The authors also generated attention maps for \ac{ad} vs. \ac{hc} vs. \ac{mci} classification for the different \textit{self-attention} heads, as well as global-level attention maps extracted from the last feature vector.

\subsubsection{Brain age regression}
Finally, \myetal{Dahan} \cite{dahan2022surface} introduced the Surface Vision Transformer, which adapted the image transformer model to surface domains. More specifically, surface meshes were transformed into triangular patches and flattened into feature vectors then inputted into the transformer model \cite{touvron2021training}.
The main task of their proposed study was to perform phenotype regression tasks using cortical surface metrics from the \ac{dhcp}.
The authors also produced average attention maps for either regression of postmenstrual age at scan and gestational age at birth.


\begin{table}[h]
\footnotesize
\centering
\tabcolsep=0.11cm
\begin{tabular}{lcccl}
\hline
\textbf{Reference} &  \textbf{Data} & \textbf{Modality} & \textbf{\#Subjects} & \textbf{Method} \\
\hline \hline \\
\multicolumn{5}{l}{\textbf{Image segmentation}} \\
\myetal{Gu} \cite{gu2020net} & In-house & sMRI (2D) & 36 & Spatial, channel and \\ 
& & & & non-local attention \\ \\
\multicolumn{5}{l}{\textbf{Disease Classification}} \\
\myetal{Jin} \cite{jin2020generalizable} & \phantom{+ }ADNI & sMRI (3D) & 1,832 & Spatial attention \\ 
& + In-house & & & \\
\myetal{Zhao} \cite{zhao2022attention} & \phantom{+ }ABIDE & fMRI (ts) & 2,622 & Time-axis attention \\ 
& + In-house & & & \\
\myetal{Sarraf} \cite{sarraf2023ovitad} & ADNI & \phantom{+ }fMRI (ts) & 1,744 & Self-attention \\ 
& & + sMRI (2D) & & \\ \\

\multicolumn{5}{l}{\textbf{Brain age regression}} \\
\myetal{Dahan} \cite{dahan2022surface} & dHCP & Surface meshes (3D) & 588 & Self-attention \\
\hline \hline
\end{tabular}
    \caption{\textbf{Articles using attention-based methods.} Abbreviations: ABIDE, Autism Brain Imaging Data Exchange; ADNI, Alzheimer’s Disease Neuroimaging Initiative; dHCP, developing Human Connectome Project; fMRI, functional Magnetic Resonance Imaging; sMRI, structural Magnetic Resonance Imaging; ts, time series.} 
    \label{tab:attention_based_methods}
\end{table}

\section{Evaluation of \ac{idl} explanations}\label{sec:evaluationofxai}

Of utmost importance, \ac{idl} explanations need to be evaluated for biological validity and robustness. Biological validity refers to whether explanations capture the true, underlying biological or pathological processes, and robustness assesses the stability of an explanation under varying conditions. Other properties of \ac{idl} explanations that were evaluated in the literature are continuity, selectivity, and downstream task performance. These properties will be discussed below in the context of the 75 studies included in this review. \\
\subsection{Biological validity}  

A key challenge for \ac{idl} in neuroimaging is that only appropriately trained medical specialists, e.g., radiologists, can validate explanations. Explanations for natural images can usually be readily validated by a general audience; for example, the model predicts ``castle" and the explanation highlights the castle turrets. In contrast, years of specialised medical training are required to identify imaging biomarkers, such as regional brain atrophy in neurodegenerative diseases. Studies may be conducted where clinicians evaluate \ac{idl} explanations. However, due to limited clinician availability, quantitative and automated validation metrics are more desirable. \\

Most of the studies we reviewed did not validate (26 out of 75) or only qualitatively validated (31 out of 75) the \ac{idl} explanations—for example, many studies compared salient brain regions identified in the explanations with those previously reported. Several fMRI studies leveraged Neurosynth, a meta-analysis platform that can return functional keywords correlated to \ac{idl} explanations, and compared these keywords against the literature.\\


The remaining 18 studies quantitatively compared \ac{idl} explanations to ground truth explanations, the latter obtained through various sources (Table~\ref{tab:evaluation_clinical_validity}). A noteworthy example is where longitudinal imaging data are available, such as in the \ac{adni} database. For subjects that progressed from \ac{cn}/\ac{mci} to \ac{ad}, a ground truth disease effect map may be computed by subtracting the registered \ac{ad} image from the \ac{cn}/\ac{mci} image \cite{baumgartner2018visual, lanfredi2020interpretation, bass2020icam, bass2022icam}. The explanation maps were then quantitatively compared to disease effect maps using \ac{ncc}. Overall, explanations from interpretable generative models achieved substantially higher \ac{ncc} (\cite{baumgartner2018visual, bass2020icam, bass2022icam}) than explanations from popular post-hoc methods, such as \textit{CAM}, \textit{Guided Backpropagation}, and \textit{Integrated Gradients}.\\

\begin{table}[!h]
\footnotesize
\centering
\tabcolsep=0.11cm
\begin{tabular}{lcccl}
\hline
\textbf{Reference} & \textbf{Ground truth data source} & \textbf{Metric} \\
\hline \hline \\
\multicolumn{3}{l}{\textbf{Perturbation-based}} \\
Liu\myetal{} \cite{liu2019deep} & cGAN-based statistics & \# brain regions \\ 
Yang \myetal{} \cite{yang2018visual} & 8 hold-out subjects & precision-recall curve \\
\\
\multicolumn{3}{l}{\textbf{Gradient-based}} \\
Levakov\myetal{} \cite{levakov2020deep} & VBM meta-analysis & Mean VBM for top 1\% regions \\ 
Ismail\myetal{} \cite{ismail2019input} & Off-task data & \% relevant features on-task \\ 
\\
\multicolumn{3}{l}{\textbf{Backpropagation-based}} \\
Thomas\myetal{} \cite{thomas2019analyzing} & NeuroSynth meta-analysis & Mean F1 score \\ 
\\
\multicolumn{3}{l}{\textbf{Weights analysis}} \\
Natekar\myetal{} \cite{natekar2020demystifying} & Ground truth segmentation & IoU \\ 
\\
\multicolumn{3}{l}{\textbf{Disentangled latent space}} \\
Mouches\myetal{} \cite{mouches2021unifying} & Ground truth segmentation & Lateral ventricle volume \\ 
Ouyang \myetal{} \cite{ouyang2022disentangling} & ADAS-Cog scores & Correlation \\
\\
\multicolumn{3}{l}{\textbf{Interpretable hybrid models}} \\
Qiu\myetal{} \cite{qiu2020development} & \textit{Post-mortem} tissue & Correlation \\ 
\\
\multicolumn{3}{l}{\textbf{Generative models}} \\
Baumgartner\myetal{} \cite{baumgartner2018visual} & ADNI disease effect map & NCC \\ 
Lanfredi\myetal{} \cite{lanfredi2020interpretation} & ADNI disease effect map & NCC \\ 
Bass\myetal{} \cite{bass2020icam} & ADNI disease effect map & NCC \\ 
Bass\myetal{} \cite{bass2022icam} & ADNI disease effect map & NCC \\ 
Sanchez\myetal{} \cite{sanchez2022healthy} & Ground truth segmentation & Dice \\
Wolleb\myetal{}\cite{wolleb2022diffusion} & Ground truth segmentation & Dice \\
Bercea\myetal{} \cite{bercea2023reversing} & Ground truth segmentation & Dice \\
\\
\multicolumn{3}{l}{\textbf{Deep structural causal models}} \\
Reinhold\myetal{} \cite{reinhold2021structural} & Image segmentation & MS lesion volume \\
\\
\multicolumn{3}{l}{\textbf{Attention}} \\
Jin\myetal{} \cite{jin2020generalizable} & AD MMSE scores & Correlation \\ 
\hline \hline
\end{tabular}
\caption{\textbf{Quantitative metrics to evaluate biological validity of \ac{idl} explanations.} Abbreviations: AD, Alzheimer’s Disease; ADAS-Cog, Alzheimer’s Disease Assessment Scale – Cognitive Subscale; ADNI,
Alzheimer’s Disease Neuroimaging Initiative; cGAN, conditional generative adversarial network; IoU, intersection over union; MMSE, Mini-mental state examination; MS, Multiple Sclerosis; NCC, normalised cross-correlation; VBM, Voxel-based morphometry.}
    \label{tab:evaluation_clinical_validity}
\end{table}
\clearpage
\subsection{Robustness}

Robustness was not evaluated in the majority of studies (62 out of 75). In the remaining studies, the robustness of population-level explanations was considered with respect to different training data (\cite{jin2020generalizable}, \cite{gao2019dense}, \cite{thibeau2020visualization}), data pre-processing methods (\cite{mellema2020architectural,li2019graph}), and model and \ac{idl} settings \cite{dvornek2019jointly,shimona2020deep, thibeau2020visualization, kim2020understanding, eitel2019testing, li2018brain, levakov2020deep}. Three studies compared population-level explanations with the same \ac{dl} task and model architecture but where the model was trained on different sources of data, and all concluded explanations were stable across datasets. For example, Jin \etal{} compared attention maps from a ResNet trained on structural \ac{mri} \ac{adni} data versus a similar in-house dataset and found the maps were significantly correlated (r=0.59) \cite{jin2020generalizable}. A few studies considered explanations trained on the same data source but with different pre-processing methodologies, investigating different atlases and atlas granularities during registration \cite{mellema2020architectural,li2019graph}. Furthermore, robustness of explanations across different model and \ac{idl} settings was evaluated, including cross-validation folds \cite{dvornek2019jointly,shimona2020deep, thibeau2020visualization, kim2020understanding}; parameter initialisation \cite{eitel2019testing, thibeau2020visualization}; hyperparameter values \cite{li2018brain,thibeau2020visualization}; and models within an ensemble \cite{levakov2020deep}. \\

Data preprocessing methods, hyperparameters, and model parameters all influence the explanations produced. Concerning data preprocessing, skull stripping often alters downstream explanations \cite{khan2019transfer, druzhinina202150}. In another example, Mellema \etal{} showed the level of atlas granularity during registration altered the important features identified for \ac{asd} classification \cite{mellema2020architectural}. \\
Concerning hyperparameters, the selection of regularisation weights for \textit{Meaningful Perturbations} changed the explanation masks for \ac{ad} classification \cite{thibeau2020visualization}. Evidence also suggests that different runs of identically trained, randomly initialised models are associated with markedly different explanations \cite{eitel2019testing, thibeau2020visualization}. It is important to be aware that bias may be present in \ac{idl} explanations from sources such as data preprocessing and hyperparameter selection and to assess explanations for such bias. \\

The robustness of explanations under different conditions may be quantitatively assessed using various similarity measures (see Table \ref{tab:evaluation_robustness}). Some studies directly compared explanations using overlap measures such as the Dice coefficient or Hausdorff distance. Other studies initially converted an explanation into a vector of mean values for $n$ atlas-derived brain regions and then compared vectors using correlation \cite{jin2020generalizable, dvornek2019jointly}, cosine similarity \cite{thibeau2020visualization} or percentage agreement between top regions \cite{eitel2019testing}. \\
\begin{table}[!h]
\footnotesize
\centering
\tabcolsep=0.11cm
\begin{tabular}{lcccl}
\hline
\textbf{Reference} & \textbf{Robustness across...} & \textbf{Metric} \\
\hline \hline \\
\multicolumn{3}{l}{\textbf{Perturbation-based}} \\
Thibeau\myetal{} \cite{thibeau2020visualization} & Datasets & Cosine similarity \\ 
Thibeau\myetal{} \cite{thibeau2020visualization} & Models (hyperparameters) & Cosine similarity \\ 
Thibeau\myetal{} \cite{thibeau2020visualization} & Models (cv folds + initialisation) & Cosine similarity \\ 
Li\myetal{} \cite{li2018brain} & Models (hyperparameters) & \# important ROIs \\ 
Eitel\myetal{} \cite{eitel2019testing}$^\dagger$ & Models (initialisation) & L2-norm \\ 
& & + relevant region coherence \\
\\
\multicolumn{3}{l}{\textbf{Gradient-based}} \\
Levakov\myetal{} \cite{levakov2020deep} & Models (ensemble) & Dice + Hausdorff distance \\ 
\\
\multicolumn{3}{l}{\textbf{Class activation maps}} \\
Kim\myetal{} \cite{kim2020understanding} & Models (cv folds) & Relevant region coherence \\ 
\\
\multicolumn{3}{l}{\textbf{Weights analysis}} \\
Dvornek\myetal{} \cite{dvornek2019jointly} & Models (cv folds) & Correlation + Dice \\ 
\\
\multicolumn{3}{l}{\textbf{Interpretable hybrid models}} \\
Shimona\myetal{} \cite{shimona2020deep} & Models (cv folds) & Mean inner-product \\ 
\\
\multicolumn{3}{l}{\textbf{Attention}} \\
Jin\myetal{} \cite{jin2020generalizable}& Datasets & Correlation \\  
\hline \hline
\end{tabular}
    \caption{\textbf{Quantitative metrics to evaluate robustness of \ac{idl} explanations.} Abbreviations: cv, cross-validation; ROI, Region of Interest. $\dagger$: Eitel \etal{} evaluated robustness across multiple method categories, not only perturbation-based.}
    \label{tab:evaluation_robustness}
\end{table}

\subsection{Other interpretable method properties}

\subsubsection{Continuity}
Similar images should have similar explanations, as originally proposed by Montavon \etal{} \cite{montavon2018methods}. Nigri \etal{} measured continuity by slightly perturbing 50 input images and then calculating the mean $L_2$-norm between explanations of the original and perturbed image \cite{nigri2020explainable}. The authors compared the continuity of the \textit{Swap Test} and \textit{Occlusion} and found the \textit{Swap Test} to be the superior method.\\

\subsubsection{Selectivity}
Regions with the highest relevance in the explanation should result in the largest change in model prediction when removed from the input image \cite{montavon2018methods}. For example, sensitivity maps were computed by Nigri \etal{}, highlighting those image regions swapped (\textit{Swap Test}) or occluded (\textit{Occlusion}) that resulted in a large change in model prediction \cite{nigri2020explainable}.  Reverse sensitivity maps were then generated by removing the complement of each image patch and recording the change in model prediction. Subsequently, Pearson correlation analysis was carried out to assess the relationship between the standard and reversed sensitivity maps, with strong negative correlations expected when the property of selectivity is satisfied. Each image from the \ac{ms} class underwent lesion in-painting, such that \ac{ms} lesions appeared to be healthy tissue in the \ac{mri} image. Explanations were generated for the original and in-painted images, and the difference between their mean values was computed, with larger differences across all images suggesting a more selective \ac{idl} method. \\

\subsubsection{Downstream task performance (disentangled latent space methods only)} Performance relates to whether the latent space distinguishes classes sufficiently for a given downstream task. In Ouyang \etal{}, \ac{dl} models were trained on the disentangled latent embeddings for two classification tasks to understand if the latent space learned a meaningful structure. The evaluation metric was the test set classification accuracy.\\

\begin{table}[!h]
\footnotesize
\centering
\tabcolsep=0.11cm
\begin{tabular}{lcccl}
\hline
\textbf{Reference} & \textbf{Interpretable method} & \textbf{Metric} \\
\hline \hline \\
\multicolumn{3}{l}{\textbf{Continuity}} \\
Nigri\myetal{} \cite{nigri2020explainable} & Perturbation-based & $L_2$ norm \\
\\
\multicolumn{3}{l}{\textbf{Selectivity}} \\
Nigri\myetal{} \cite{nigri2020explainable} & Perturbation-based & Correlation\\
\\
\multicolumn{3}{l}{\textbf{Downstream task performance (disentangled latent space methods only)}} \\
Ouyang\myetal{} \cite{ouyang2022disentangling} & Disentangled latent space & Classification accuracy \\
\hline \hline
\end{tabular}
    \caption{\textbf{Other properties and quantitative metrics for iDL explanations.}}
    \label{tab:evaluation_other_properties}
\end{table}

\section{Discussion and conclusion}\label{sec:discussion}

In this review, we identified 75 neuroimaging studies that utilised \ac{idl} methods, and we classified the methods into five post-hoc and five intrinsic categories. To the best of our knowledge, this is the first systematic review of \ac{idl} in neuroimaging with a notably more extensive review of intrinsic methods than found in the literature \cite{thibeau2023interpretability}. In addition, we found five properties of \ac{idl} explanations that were investigated and are important when considering the suitability of an \ac{idl} method for adoption. \\

The most common \ac{idl} methods utilised were \textit{class activation maps}, \textit{perturbation-based} and \textit{gradient-based} methods. Post-hoc methods are popular because they are well-established in computer vision tasks, easy to implement, and readily available in \ac{dl} packages. However, historically, post-hoc methods were designed for and validated on natural images and may be inappropriate for neuroimaging tasks. For example, saliency methods were shown to only focus on a few discriminative features of a class \cite{zhang2021explainable, bass2020icam}, rather than identifying all imaging features, which may be important for diagnosis and treatment. Their reliability is also questionable as some post-hoc methods, in particular \textit{Guided backpropagation} and \textit{Guided Grad-CAM} \cite{adebayo2018sanity}, still produce convincing explanations despite randomised model weights or data labels. In contrast, intrinsic methods are generally more appropriate for neuroimaging because they are designed specifically for the application \textit{e.g.}, constructing a causal graph specific to \ac{ms} \cite{reinhold2021structural}. Additionally, generative models produced explanations with substantially higher correlation to ground-truth disease markers compared to explanations from several post-hoc methods \cite{baumgartner2018visual, bass2020icam, bass2022icam}. Nevertheless, intrinsic interpretable deep learning is still an emerging field, and such methods are currently more time-consuming to implement than post-hoc methods. \\

We will now provide some recommendations for researchers when using \ac{idl} with neuroimaging datasets. First, we suggest utilising multiple \ac{idl} methods, including several across different post-hoc method categories (such as \textit{Occlusion}, \textit{LRP}, and \textit{GradCAM}) and one intrinsic method that is best suited for the project application, end-user requirements, objectives, etc. It is important to carefully select one intrinsic method during the design phase as it can be time-consuming to implement. For pre-existing models, incorporating an appropriate attention mechanism and retraining the model may be feasible. Then, compare explanations from different methods and prioritise features highlighted across all methods. \\
Second, recall that various confounding factors, such as data preprocessing, random initialisation, and cross-validation, can affect explanations. Therefore, we advise averaging explanations across cross-validation folds and multiple runs to improve robustness. Also, consider visualising explanations for a reasonable selection of model preprocessing and hyperparameter settings. If using multiple neuroimaging datasets, we recommend adopting a standardised pre-processing pipeline to reduce the risk of biased explanations. \\

Third, validating explanations across an entire test dataset is crucial rather than limiting assessments to a select few samples. This comprehensive validation helps ensure the generalisability of the explanations. Consider acquiring ground truth to validate explanations quantitatively, such as computing disease affect maps from longitudinal imaging datasets. If not possible, then impartially compare explanations to existing physiopathological literature. In summary, do not unquestioningly trust the explanations produced by an \ac{idl} method. \\

When applying \ac{idl} methods for neuroimaging, an important concern is the complexity of the biological mechanisms underlying the data and the interactions between multiple imaging features. Many interpretability methods identified in this review do not consider the causal mechanisms that contribute to the data nor the impact of confounding factors in the explanations. We have, however, discussed state-of-the-art causal models that attempt to address causality in interpretability, and we foresee such models will play an important role in the future of \ac{idl} \cite{pawlowski2020deep, reinhold2021structural}. We also conclude a suite of standardised, quantitative evaluation metrics to compare performance across \ac{idl} methods needs to be established to promote the trustworthiness of \ac{idl} methods. \\  

\section{Acknowledgments}

We would like to thank the EPSRC Centre for Doctoral Training in Smart Medical Imaging [EP/S022104/1] for funding studentships for LM and MDS. Thanks also to the National Brain Mapping Laboratory (NBML) and the IUMS Medical Physics Department for supporting and financing the Masters program for FH.

The images used in some of the figures in this work were funded by the Alzheimer’s Disease Neuroimaging Initiative (ADNI) [National Institutes of Health Grant U01 AG024904] and DOD ADNI [Department of Defense award number W81XWH-12-2-0012].

Declarations of interest: The authors have no conflicts of interest to disclose.

\bibliographystyle{splncs04}
\bibliography{references}

\clearpage
\printacronyms[include=abbrev,name=Glossary]

\appendix

\section*{Appendix}
 
\subsection*{Integrated Gradients}
\label{appendix:A}
The Integrated Gradients equation is:

\begin{equation}
\text{\textit{Integrated Gradients}} \, (I_0) = (I_0 - I_b) \times \frac{1}{m} \sum_{k=1}^m \frac{\partial S_c(I_b + \frac{k}{m} (I_0 - I_b) )}{\partial I}
\end{equation}
 
\subsection*{Layer-wise Relevance Propagation (LRP)}
\label{appendix:B}
\textbf{\textit{\Ac{lrp}-0}} is the most basic relevance propagation rule, defined as (Eq.~\ref{eq:lrp0}):

\begin{equation}\label{eq:lrp0}
R_{i}^{l} =\sum_{j}\frac{z_{i j}}{z_{j}} R_{j}^{l+1}
\end{equation} 

where $R_{i}^{l}$ is the relevance value for node $i$ at layer $l$, while $z_{ij}=a_{i}w_{ij}$ and $z_{j} = \sum_{i} z_{ij}$. Here, $a_{i}$ is the activation function for node $i$, while $w_{ij}$ is the weight between node $i$ in layer $l$ and node $j$ in layer $l+1$. \\

\textit{\textbf{\ac{lrp}-$\epsilon$}} produces sparser relevance maps by introducing a small, positive $\epsilon$ term in the denominator (Eq.~\ref{eq:lrp_epsilon}): 

\begin{equation}\label{eq:lrp_epsilon}
      R_{i}^{l} =\sum_{j} \frac{z_{ij}}{z_{j} + \epsilon} R_{j}^{l+1}
\end{equation}
\\

\textit{\textbf{\ac{lrp}-$\alpha\beta$}} treats positive weights (indicated by $+$) and negative weights (indicated by $-$) separately (Eq.~\ref{eq:lrp_alphabeta}) and weights their contribution with hyperparameters $\alpha$, $\beta$ respectively (with constraints $\alpha - \beta = 1, \beta \ge 0$):

\begin{equation}\label{eq:lrp_alphabeta}
   R_{i}^{l} =\sum_{j}(\alpha \frac{z_{ij}^{+}}{z_{ j}^{+}}-\beta \frac{z_{ij}^{-}}{z_{j}^{-}}) R_{j}^{l+1}
\end{equation}

\end{document}


\maketitle

\section{Systematic Search Methodology}

We identified relevant articles by querying PubMed, Web of Science, Google Scholar and arXiv using specific search terms. The search terms used are listed in Fig. \ref{search_terms}. All fields in the database were queried, with the exception of Google Scholar where full texts were searched instead. Articles from 2015 onwards only were included for PubMed and Google Scholar, whereas all years were included for Web of Science and arXiV due to the small number of articles returned. \\

Next we screened articles based on article title and abstract. We formulated a set of inclusion and exclusion criteria and accepted or rejected articles based on these criteria. The screening criteria are listed in Table \ref{tab:screening_criteria}. Only the first 500 results from Google Scholar were screened because later results were largely irrelevant. \\

\begin{table}[]
    \centering
    \begin{tabular}{ p{12cm} }
        \hline
        \textbf{Inclusion/ exclusion criteria for article screening} \\ 
        \hline
        \textbf{Include}...both \emph{in-vivo} and \emph{ex-vivo} imaging.  \\
        \textbf{Exclude}...non-human subjects. \\
        \textbf{Include}...the following imaging modalities: structural and functional MRI, CT, PET, DWI/tractography. \\
        \textbf{Exclude}...EEG and MEG data. \\
        \textbf{Include}...both peer reviewed and non-peer reviewed articles. \\
        \textbf{Exclude}...non-English language articles. \\
        \textbf{Exclude}...PhD and Masters theses. \\
        \textbf{Exclude}...reviews, surveys, opinion articles and books. Articles must implement at least one interpretable deep learning method. \\
        \textbf{Exclude}...interpretable methods applied to machine learning models other than neural networks. For example, decision trees, random forests, SVMs, Gaussian processes. \\
        \textbf{Exclude}...for quality control. For example, some methods claimed to be interpretable but were not. \\
        \hline
    \end{tabular}
    \caption{Inclusion and exclusion criteria for title and abstract screening}
    \label{tab:screening_criteria}
\end{table}

After screening, we extracted data that were relevant to our review questions from all accepted articles into a table. We extracted 27 data points covering 6 different topics: article, imaging, modelling, interpretability method, interpretability method evaluation and study limitations (see Table \ref{tab:article_data_collection} in appendix).

\begin{figure}
     \centering
     \begin{subfigure}[b]{0.4\textwidth}
         \centering
         \includegraphics[width=7cm]{images/studies_by_year.jpg}
         \caption{Number of studies by year}
         \label{fig:studies_by_year}
     \end{subfigure}
     \hfill
     \begin{subfigure}[b]{0.4\textwidth}
         \centering
         \includegraphics[width=7cm]{images/studies_by_dataset.jpg}
         \caption{Number of studies by dataset. TO ADD ACRONYMS.}
         \label{fig:studies_by_dataset}
     \end{subfigure}
     \hfill
     \begin{subfigure}[b]{0.4\textwidth}
         \centering
         \includegraphics[width=7cm]{images/studies_by_modality.jpg}
         \caption{Number of studies by imaging modality.}
         \label{fig:studies_by_modality}
     \end{subfigure}
    \caption{Number of studies across years, datasets and imaging modalities.}
    \label{fig:studies_by_year_dataset_modality}
\end{figure}

The count of neuroimaging studies applying interpretable deep learning methods have approximately doubled annually\footnote{note, the cutoff date of this review was part way through 2021} (Fig. \ref{fig:studies_by_year}). Most studies used existing public medical image datasets (76\%), with the most popular being the Alzheimer's Disease Neuroimaging Initiative (ADNI, 37\% of studies) followed by the Human Connectome Project (HCP, 17\% of studies) (Fig. \ref{fig:studies_by_dataset}). The majority of studies (90\%) are either structural or functional magnetic resonance imaging (MRI). \\

\begin{figure}
     \centering
     \begin{subfigure}[b]{0.4\textwidth}
         \centering
         \includegraphics[width=7cm]{images/studies_by_clinical_task.jpg}
         \caption{Number of studies by clinical task}
         \label{fig:studies_by_task}
     \end{subfigure}
     \hfill
     \begin{subfigure}[b]{0.4\textwidth}
         \centering
         \includegraphics[width=7cm]{images/studies_by_method.jpg}
         \caption{Number of studies by interpretable method}
         \label{fig:studies_by_method}
     \end{subfigure}
        \caption{Number of studies across clinical tasks and interpretable methods}
        \label{fig:studies_by_task_and_method}
\end{figure}